\documentclass[a4paper,twocolumn,fleqn]{article}

\usepackage{amsmath}             
\usepackage{txfonts}               
\usepackage{bm}                  
\usepackage{cite}                  
\usepackage{graphicx}
\usepackage{pfr}                   

\usepackage[T1]{fontenc} 

\usepackage[normalem]{ulem} 
\usepackage{units} 
\usepackage{float} 
\usepackage{hyperref}

\AtBeginDocument{\mathcode`v=\varv} 

\usepackage{dcolumn}
\hypersetup{
    colorlinks=true,
    linkcolor=blue,
    urlcolor=blue,
    citecolor=blue,
    pdfpagemode=FullScreen,
    }

\AtBeginDocument{%

}

\newcommand{\Btheta}{\dot{B}_\mathrm{\theta}}
\newcommand{\Te}{\tilde T_\mathrm{e}}

\newcommand{\Core}{\mathrm{Core}}
\newcommand{\Inter}{\mathrm{Inter}}
\newcommand{\Edge}{\mathrm{Edge}}
\newcommand{\betaN}{\beta_{\mathrm{N}}}
\newcommand{\qedge}{q_{\mathrm{95}}}
\newcommand{\Env}{\mathrm{Env}}

\newcommand{\ConAvgNormAmpEnv}{\left< A^\textrm{Env}/A^\textrm{Env}_\textrm{max} \right>}

\usepackage[utf8]{inputenc}
\usepackage[english]{babel}

\begin{document}

\title{\vspace{-0.6cm}Study of Low-Frequency Core-Edge Coupling in a Tokamak: \\ I. Experimental Observation in KSTAR}

\author{Wonjun~LEE\sup{1,2}, Andreas~BIERWAGE\sup{3}, Seungmin~BONG\sup{1}, Jaewook~KIM\sup{4}, K.D.~LEE\sup{4}, J.G.~BAK\sup{4}, G.J.~CHOI\sup{1}, C.~SUNG\sup{1}, Y.-c. GHIM\sup{1\ast}}
\affiliation{
	\sup{1} Department of Nuclear and Quantum Engineering, KAIST, Daejeon 34141, Republic of Korea \\
	\sup{2} Max-Planck-Institut f\"ur Plasmaphysik, Wendelsteinstrasse 1, D-17491 Greifswald, Germany \\
	\sup{3} National Institutes for Quantum Science and Technology (QST), Rokkasho Institute for Fusion Energy, Aomori 039-3212, Japan\\
	\sup{4} Korea Institute of Fusion Energy, Daejeon 34133, Republic of Korea
}

\date{}

\email{ycghim@kaist.ac.kr}

\begin{abstract}
Double-peaked fishbone events across multiple KSTAR discharges are investigated. The normalized beta $\betaN$ and the edge safety factor $\qedge$ under which the fishbones appear vary depending on the presence and form of external magnetic perturbations. The fishbone strength is closely related to $\betaN$ and $\qedge$: as $\betaN$ increases and $\qedge$ decreases, the fishbone strength increases. Measured fishbone-relevant signals are decomposed into amplitude envelope and phase components in the temporal domain, which are analyzed separately. In terms of the amplitude envelope component, the edge electron temperature fluctuation $\Te^{\Edge}$ becomes more correlated with the poloidal magnetic fluctuation $\Btheta$ compared to the core electron temperature fluctuation $\Te^{\Core}$ as fishbone strength increases. In terms of the phase component, the phase of $\Te^{\Edge}$ precedes the phase of $\Te^{\Core}$ except in the case of very weak fishbones where the phase relations are inconclusive due to weak fishbone activity at the edge plasma, which is comparable to background fluctuations. The investigation suggests the possibility that the edge activity is not a mere side effect of the core activity, but could play an active role.
\end{abstract}


\maketitle

\thispagestyle{empty} 
\pagestyle{plain} 
\everypar{\looseness=-1} 

\tableofcontents

\section{Introduction}

The fishbone instability is a low-frequency type of energetic particle mode (EPM) that usually occurs in the form of repeated burst of electromagnetic fluctuations that exhibit downward frequency chirping on the time scale of a few milliseconds or less. It is associated with a redistribution or loss of energetic particles (EP) such as beam ions. Since its first discovery in PDX in the 1980s~\cite{PDX_1983}, similar fishbone-like modes have been routinely observed in various tokamaks~\cite{Heidbrink_1990, Nave_1991, Kass_1998, EAST_2015}. This ``classical'' PDX-type fishbone instability occurs when the tokamak safety factor $q$ in the plasma core is sufficiently close to or below unity ($q\sim1$). Its dominant poloidal ($m$) and toroidal ($n$) Fourier mode numbers are $(m, n)=(1, 1)$, but $m\ge2$ harmonics can also be significant and contribute to the transport of EPs during the course of a fishbone burst~\cite{PDX_1983,White_1983}.

When the bulk plasma is modeled in the ideal MHD limit, the underlying mode can be viewed as a resonant branch of a marginally stable internal kink that emerges under the influence of and is destabilized by trapped EPs whose precession frequency is comparable to the mode frequency (usually around $10$~kHz or less in the plasma frame)~\cite{Chen_1984}. This branch is sometimes called ``precessional fishbone.'' When finite-Larmor-radius effects play a significant role, a marginally stable branch of drift-Alfv\'{e}n waves can also be destabilized by particles that resonate with the ion diamagnetic drift frequency~\cite{Coppi_1986}. This branch tends to have a lower frequency and is sometimes called ``diamagnetic fishbone.'' Nonperturbative EP and diamagnetic effects can also contribute simultaneously~\cite{Zhang1989}.

A comprehensive theoretical framework including non-linear multi-scale dynamics has been presented by Zonca, Chen and co-workers around 2015~\cite{Zonca_2015, Chen2016}.  Due to the fluctuating fields, resonant EPs experience cumulative radial transport over the course of many bounce or transit periods in the plasma's nonuniform equilibrium magnetic field.  Wave-EP phase locking plays a crucial role by counteracting resonance detuning, thereby broadening the resonance region and prolonging the resonant drive. As EPs are transported radially (down their density gradient), their transit frequency decreases. In the case of precessional fishbones, this can explain the rapid (sub-milliseconds time scale) downward chirp of the mode frequency via the self-consistent coupling between the mode and the EP dynamics.

Meanwhile, around 2009, another type of mode with fishbone-like behavior has been observed in JT-60U~\cite{Matsunaga2009} and DIII-D experiments~\cite{Okabayashi2009} (e.g., see ~\cite{Matsunaga_2013} and references therein). In contrast to the core-localized classical fishbone, this mode tends to peak near the $q=2$ surface and appears to couple to the non-ideal wall, so it has thus been dubbed ``off-axis fishbone'' or ``energetic particle-driven wall-stabilized mode.'' Like the classical fishbone, the off-axis mode is also dominated by the toroidal Fourier mode number $n=1$, albeit its characteristic waveform distortion seen in magnetic and density fluctuations~\cite{Matsunaga_2013} implies the presence of higher harmonics ($n\ge2$)~\cite{Li2023}.

Except for possible minor radial amplitude modulations, the radial mode structure of both the classical PDX-type fishbone~\cite{PDX_1983, Heidbrink_1990, Nave_1991, Kass_1998, EAST_2015} as well as the off-axis fishbone~\cite{Matsunaga2009, Okabayashi2009, Matsunaga_2013, Li2023} is assumed to have only one peak, and in conventional tokamaks both seem to be driven by trapped energetic ions whose orbits are populated by transverse beam injection.

Recently, yet another type of fishbone-like instability has been discovered in KSTAR experiments~\cite{Wonjun_2023}. This double-peaked fishbone mode features one peak in the core and another in the edge region of the plasma, with a distinct lack (or deep minimum) of activity in the intermediate region. Both peaks burst and chirp in unison, in spite of the fact that these plasmas are subject to differential toroidal rotation due to dominant unbalanced tangential beam injection, which also implies that the population of trapped ions is relatively small in the high-energy tail. The explanation of this mode still poses a challenge and includes questions like: (i) how is this mode destabilized?, and (ii) how are the oscillations at the inner and outer peak synchronized with respect to both amplitude and phase in a differentially rotating plasma?

With the purpose of providing more information that could help to understand and, ultimately, control or even utilize the double-peaked fishbone phenomenon, the present paper reports additional data based on experimental measurements and comprehensive statistical analyses. The specific path chosen for this work was motivated by a curious question that is relevant to both the problem of drive (i) and radial coupling (ii): could it be that the fluctuations near the magnetic axis constitute a parasitic secondary mode that is excited (via some yet-to-be identified coupling mechanism) by the outer mode component near the plasma edge?

This idea, in turn, is based on several thoughts and existing pieces of evidence: 
\begin{itemize}
\item Fast ion losses with strong correlation with the fishbone burst (including chirping patterns) were detected at multiple divertor Langmuir probes (to be reported in detail elsewhere). This indicates that resonant interactions occur in the edge region.
\item The plasma at hand are primarily driven by tangential beams. This reduces the likelihood for the two peaks being jointly driven by barely trapped energetic ions on wide drift orbits that traverse both the core and the edge, as we had tentatively speculated in ~\cite{Wonjun_2023}. Therefore, it is necessary to consider the possibility of radially localized drive, which would imply that only one peak may be driven by resonant EPs. (Resonance analyses and numerical computations of the beam ion distribution function are underway and will be presented in a future paper of this series.)
\item Although the beam ions are primarily deposited in the inner core, the plasmas at hand are subject to sawtooth crashes, which may increase the population of fast ions in the outer core, especially the population of passing fast ions~\cite{Kolesnichenko1996}. (Moreover, note that, in principle, there also exists the possibility of thermal ion drive~\cite{Du_2021, Liu2022}, especially in H-mode plasmas, which have steep edge gradients.
\item Unless the mode's outer component carries much less energy than the inner component\footnote{Unfortunately, we are currently unable to determine the absolute local amplitude of the fluctuations from ECE measurements at different radii, so it is difficult to reliably estimate the relative energy contents of the fishbone's inner and outer components.}, inward drive would also be favored geometrically by volumetric focusing (towards the $1/r$ singularity at the magnetic axis with minor radius $r=0$).  
\end{itemize}
Numerical studies discussing these and other possibilities will be reported in subsequent papers of this series, beginning with Ref.~\cite{Bierwage2026}.

Thus motivated, the properties of thousands of double-peaked fishbones across dozens of KSTAR discharges are thoroughly examined, looking at representative individual events as well as statistics. In Section \ref{sec:stat_DPFB} the plasma conditions in which double-peaked fishbones appear are reported. The fishbone activity signals are decomposed into amplitude envelope and phase components in the temporal domain and analyzed separately for multiple shots in Section \ref{sec:Experimental_observation}. In Section \ref{sec:anal_method} the data analysis methods for amplitude envelope and phase components are described and the corresponding results are presented in Section \ref{sec:results}. Our work is summarized in Section \ref{sec:summary}.

\section{Observational conditions for the double-peaked fishbones in KSTAR}\label{sec:stat_DPFB}

The double-peaked fishbones appear in H-mode plasmas with and without external magnetic perturbations~\cite{Park2024} in KSTAR, but that does not mean that the characteristics of the fishbones are identical regardless of the presence of those perturbations. We have surveyed 40 KSTAR discharges and around 3,000 double-peaked fishbones, and analyzed the distribution of fishbone strength. Hereafter, the attribute ``double-peaked'' is usually omitted, unless needed for emphasis or distinction. In this work, fishbone strength refers to the maximum magnitude of the poloidal magnetic fluctuation $|\Btheta|_{\mathrm{max}}$ measured by a Mirnov coil at the plasma facing component near the outer midplane, filtered in the frequency range of the fishbone, during the lifetime of each fishbone. One should thus bear in mind that the Mirnov coil signal amplitude may represent mainly the strength of the nearer (outer) component of a double-peaked fishbone than its more distant (inner) peak, unless the outer peak is too weak.

\begin{figure}
	[tb]\centering\vspace{-0.3cm}
	\includegraphics[width=0.45\textwidth,keepaspectratio]{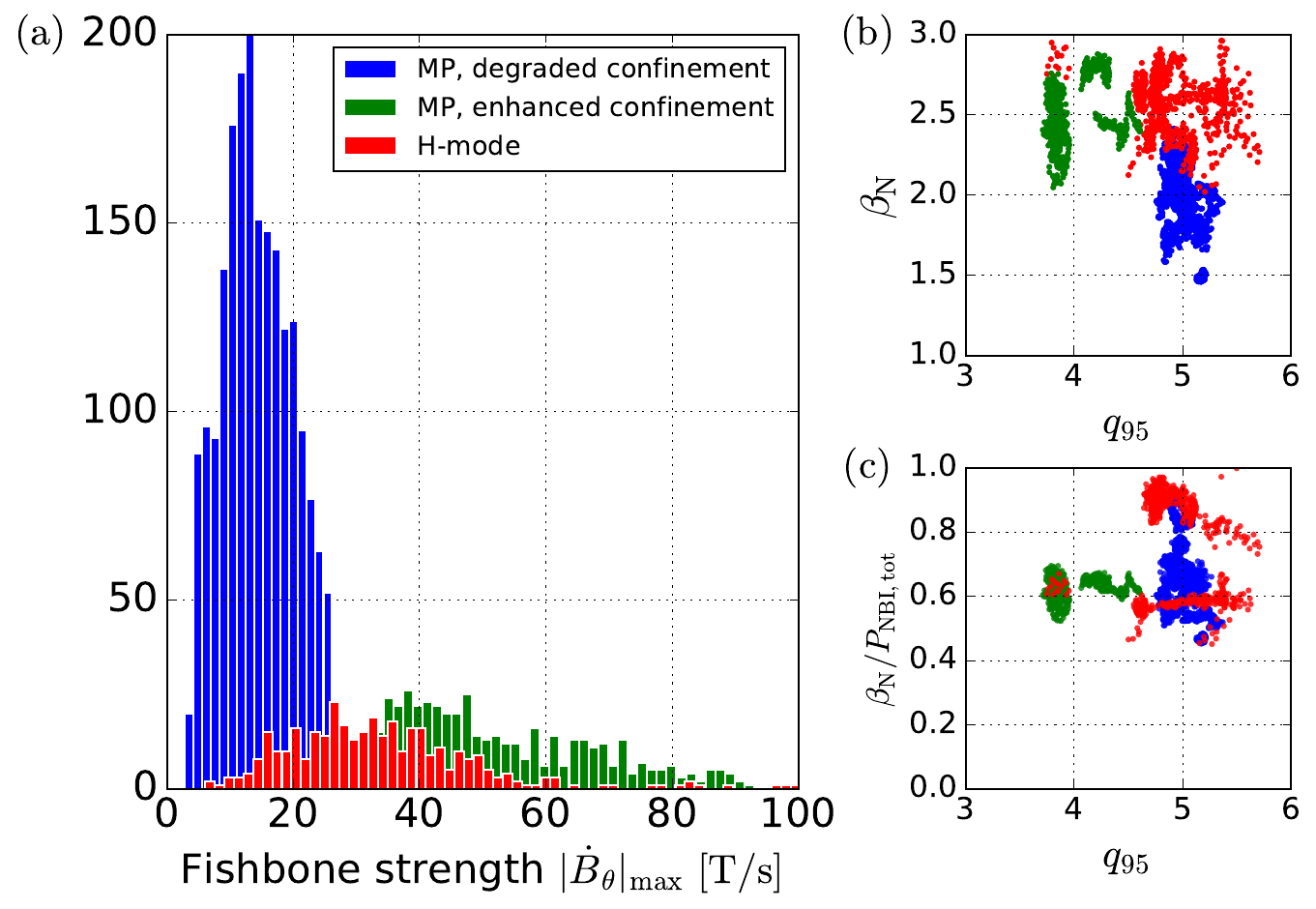}\vspace{-0.2cm}
	\caption{(a) Histograms of double-peaked fishbone strength in KSTAR. 40 KSTAR discharges and around 3,000 fishbones are examined. Blue and green bars are distributions of the fishbone strengths found in H-mode plasmas with magnetic perturbations that cause energy confinement degradation (blue) and enhancement (green). Red bars show the distribution of the fishbone strengths in H-mode plasmas without external magnetic perturbations. (b) The edge safety factor at the 95\% normalized flux surface $\qedge$ versus the normalized beta $\betaN$ when the fishbones are observed. Red, blue, and green dots correspond to the same color code as in (a). (c) Same as in (b) except the ordinate is $\betaN$ divided by the total NBI power $P_\mathrm{NBI, tot}$.}\vspace{-0.3cm}
	\label{fig:Stat_DBFB}
\end{figure}

In Figure~\ref{fig:Stat_DBFB}(a), histograms of fishbone strength are presented. The blue histogram shows the distribution of fishbone strength that appears in H-mode plasmas with ordinary magnetic perturbations. Here, ordinary magnetic perturbations refer to those that cause a density pump-out and energy confinement degradation. Typically, weak fishbones whose strength lies below 20~T/s are observed in the presence of ordinary magnetic perturbations. Modest strength fishbones, $20\lesssim |\Btheta|_{\mathrm{max}} \lesssim 40$~T/s, appear in H-mode plasmas without external magnetic perturbation, and are shown with the red bars in Figure~\ref{fig:Stat_DBFB}(a). Strong fishbones, $|\Btheta|_{\mathrm{max}}\gtrsim 40$~T/s, appear in plasmas with optimized magnetic perturbations that enhance energy confinement via the magnetic braking effect~\cite{Kimin_2023}, and are represented by the green bars. Let us emphasize once more that, as the magnetic fluctuation is measured outside the plasma, this classification is likely to represent the strength of the mode's edge component, which is closer to the Mirnov coil.

The injected neutral beam (NBI) power\cite{BAE20121597} associated with the blue histograms in Figure \ref{fig:Stat_DBFB}(a) ranges from $2.5$~MW to $3.8$~MW. For the red histograms, the NBI power ranges from $2.5$~MW to $4.5$MW, while for the green histograms, it ranges from $3.8$~MW to $4.3$~MW. 

The edge safety factor at the 95\% normalized flux surface $\qedge$ and the normalized beta $\betaN$ during fishbone events are estimated. The result is shown in Figure \ref{fig:Stat_DBFB}(b). Red, blue, and green dots correspond to the same color code as in Figure \ref{fig:Stat_DBFB}(a). Not only do fishbone strengths differ, but their $\qedge$ and $\betaN$ also vary depending on the presence and form of external magnetic perturbations. Figure \ref{fig:Stat_DBFB}(c) shows a reorganized plot of panel (b) where the ordinate is the $\betaN$ normalized by the total NBI power. Vertical scatters shown in Figure \ref{fig:Stat_DBFB}(b) become better grouped in panel (c), and we observe that the red dots are separated into two groups. This is associated with the KSTAR NBI power being discrete using three separate and independent beam lines. The lower(upper) group is from KSTAR discharges using three(two) beam lines.

In H-mode plasmas with ordinary magnetic perturbations (blue dots), causing density pump-out and energy confinement degradation, the fishbones appear in the lowest normalized beta $\betaN$ and high edge safety factor $\qedge$ among the three cases, i.e. $1.5\lesssim\betaN\lesssim 2.3$ and $4.5\lesssim \qedge \lesssim 5.5$. Without magnetic perturbations (red dots) the fishbones appear in the relatively high $\betaN$ and high $\qedge$, i.e. $2\lesssim\betaN\lesssim 3$ and $4.5\lesssim \qedge \lesssim 5.5$. Note that a small portion of these fishbones have smaller $\qedge$ around 3.8. With optimized magnetic perturbations, causing energy confinement enhancement (green dots), strong fishbones appear in the high $\betaN$ and the lowest $\qedge$, i.e. $2\lesssim\betaN\lesssim 3$ and $3.7\lesssim \qedge \lesssim 4.5$. As $\qedge$ decreases and $\betaN$ increases, we observe that the fishbone strength $|\Btheta|_{\mathrm{max}}$ increases.

\section{Analysis of amplitude envelope and phase components of the double-peaked fishbone activity} \label{sec:Experimental_observation}

Fluctuating signals of fishbone activities can be decomposed into amplitude envelope and phase components in the temporal domain. In this section, the amplitude envelope ($A^\textrm{Env}$) and phase ($\phi$) components are separately analyzed. The instantaneous amplitude envelope and phase are obtained using the Hilbert transform. Electron cyclotron emission images~\cite{KSTAR_ECEI_2014} of the fishbones are also presented if they are available (see Section \ref{sec:investigated_shots}).

\subsection{Data analysis method}
\label{sec:anal_method}

\subsubsection{Amplitude envelope component}\label{sec:method_env}

Amplitude envelopes ($A^\textrm{Env}$) of fluctuating temperature $\Te$, measured by the ECE (electron cyclotron emission) radiometer~\cite{KSTAR_ECE}, are conditionally averaged~\cite{Block_2006}, and the result characterizes the average temporal structure of $\Te$ along radially located ECE channels. Signals from a Mirnov coil~\cite{KSTAR_Mirnov}, which measures the poloidal magnetic fluctuation $\Btheta$ and from the ECE radiometer are filtered within the frequency range of the fishbones and the amplitude envelopes of the filtered signals are estimated. All demonstrated ECE channels are valid local fluctuation measurements.

\begin{figure}
    [tb]\centering\vspace{-0.55cm}
    \includegraphics[width=0.35\textwidth,keepaspectratio]{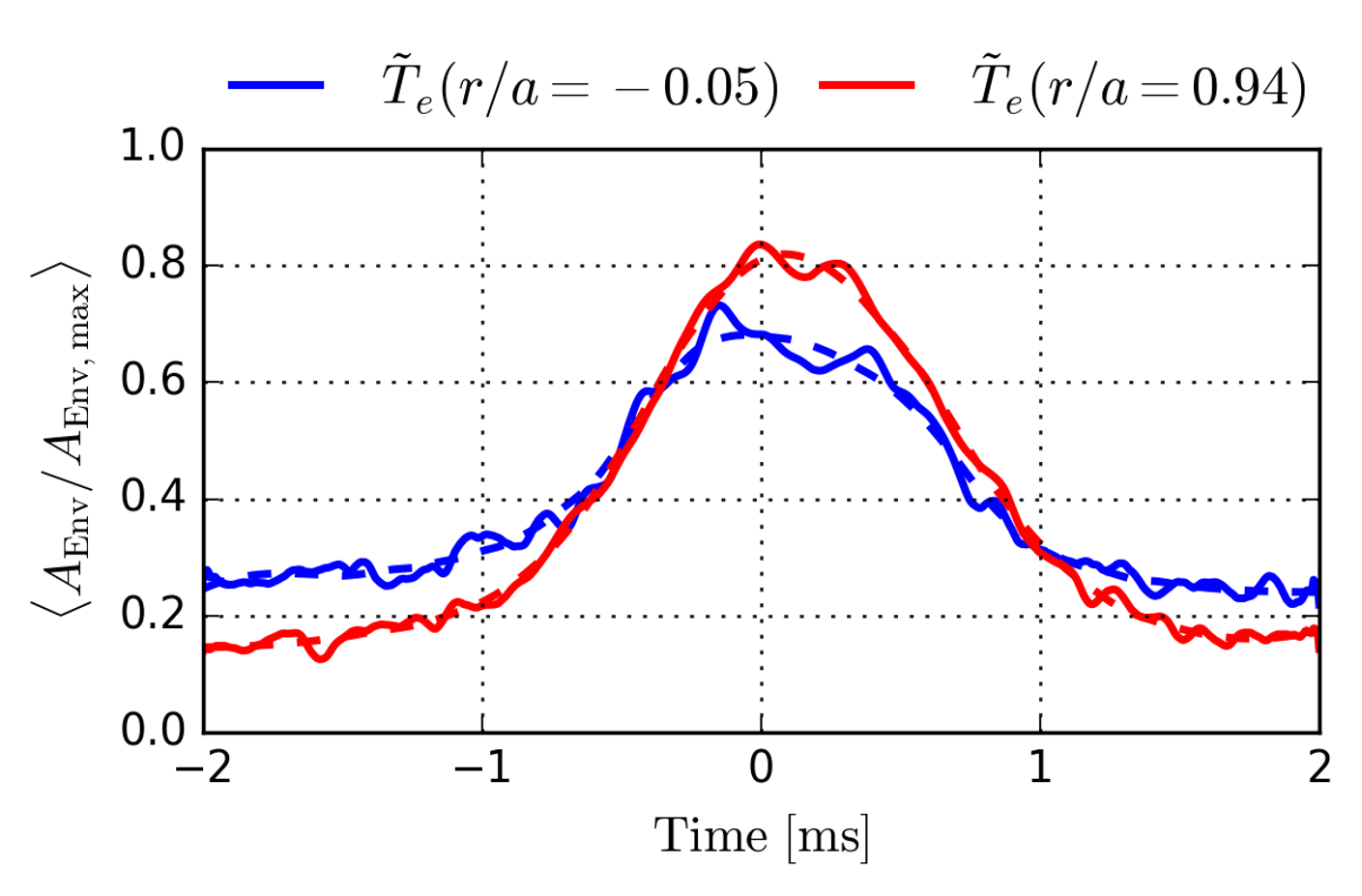}\vspace{-0.2cm}
    \caption{Examples of the conditional averages of normalized amplitude envelopes $\ConAvgNormAmpEnv$ (solid lines) of fishbone-relevant $\Te$ measured by two ECE channels, i.e., $r/a=-0.05$ (blue) and $r/a=0.94$ (red). The dashed lines are conditional averages smoothed by the Savitzky-Golay filter.}\vspace{-0.3cm}
    \label{fig:smoothing_cond_avg}
\end{figure}

The signals from the  \textit{i}$^\textit{th}$ fishbone in the time range $t_{i,\mathrm{ref}}- \Delta t_{1} \leq t_{i} \leq t_{i,\mathrm{ref}}+ \Delta t_{2}$ are collected for the conditional average. The reference time of the \textit{i}$^\textit{th}$ fishbone, $t_{i,\mathrm{ref}}$, is the time at which the amplitude envelope of $\Btheta$ reaches its maximum value, i.e. $t_{i,\mathrm{ref}}=t_{i}(|\Btheta|_{i}=|\Btheta|_{i,\mathrm{max}})$. $\Delta t_{1}$ and $\Delta t_{2}$ are of the order of a milli-second, determined by the average growth and decay times of the analyzed fishbones. For instance, in Figure \ref{fig:smoothing_cond_avg}, $\Delta t_{1}=\Delta t_{2}=2$~ms. Each collected amplitude envelope is normalized by its maximum amplitude before the conditional average. Although normalization is not required for the conditional average, we do so to focus on the relative temporal behavior of the amplitude envelopes.

The conditional average of normalized amplitude envelopes (denoted as $\ConAvgNormAmpEnv$) for $\Btheta$ has a value of one at its peak by definition because the reference time is when the amplitude envelope of $\Btheta$ reaches its maximum value. $\ConAvgNormAmpEnv$ for $\Te$ will range between zero and unity. The closer the $\ConAvgNormAmpEnv$ for $\Te$ gets to unity, the more consistent the time difference is between peaks of the amplitude envelopes ($A^\textrm{Env}$) of $\Te$ and $\Btheta$ across all fishbone events. 

Collected amplitude envelopes of the fishbones are used to estimate the correlation coefficient between $\Btheta$ and $\Te$ during the fishbone burst which demonstrates a measure of the correlation along the radial ECE channels, while $\ConAvgNormAmpEnv$ focuses more on the average temporal structure relative to $\Btheta$. We estimate the correlation coefficient as: 
\begin{equation}
\label{eq:corr}
C_{\Btheta \Te}\equiv\frac{\langle (\Btheta(t) -\mu_{\Btheta})(\Te(t) - \mu_{\Te}) \rangle}{\sqrt{ \langle (\Btheta - \mu_{\Btheta})^{2} \rangle\langle (\Te - \mu_{\Te})^{2} \rangle }},
\end{equation}
where $\langle \cdot \rangle$ is an ensemble average over multiple fishbones and $\mu_{X}$ denotes the mean of signal $X$.
A standard deviation $\sigma_{\Btheta \Te}$ of the correlation coefficient is determined by:
\begin{equation}
\label{eq:corr_sd}
    \sigma_{\Btheta \Te}^2 \equiv \frac{\langle (c_{\Btheta \Te} - \mu_{c})^{2} \rangle}{\langle (\Btheta - \mu_{\Btheta})^{2} \rangle\langle (\Te - \mu_{\Te})^{2} \rangle},
\end{equation}
where $c_{\Btheta \Te}= (\Btheta(t) -\mu_{\Btheta})(\Te(t) - \mu_{\Te})$ and $\mu_{c}= \langle (\Btheta(t) -\mu_{\Btheta})(\Te(t) - \mu_{\Te}) \rangle$.

Differentials of $\ConAvgNormAmpEnv$, denoted as $\Delta_{t}\ConAvgNormAmpEnv$, are also estimated to indicate the average growth and decay intervals of the amplitude envelopes. Here, $\Delta_{t}$ denotes a finite difference in time domain. If either the core or edge activity of the fishbone starts to grow earlier than the other, the differentials show such a feature. Before performing the differentials, $\ConAvgNormAmpEnv$ is smoothed by the Savitzky-Golay filter. 

In Figure \ref{fig:smoothing_cond_avg}, examples of the conditional averages of normalized amplitude envelopes (solid lines) and their smoothed ones (dashed lines) are shown for $\Te$ measured by two ECE channels at $r/a=-0.05$ (blue) and $r/a=0.94$ (red). Here, $r/a$ is the normalized radial position of the ECE channel calculated as $(R-R_\mathrm{axis})/a$ with the major radius position $R$ of the ECE channel (based on the magnitude of the total magnetic fields) at the midplane, major radius position of the magnetic axis $R_\mathrm{axis}$ and the minor radius $a$ obtained from the EFIT. Since we use 12 radial channels of the ECE radiometer, $\ConAvgNormAmpEnv$ and $\Delta_{t}\ConAvgNormAmpEnv$ are also visualized in 2D contours in this work.

\subsubsection{Phase component}

In terms of the phase ($\phi$) component, we examine if there exist any fixed phase relationships between the core and the edge fluctuations of electron temperature during fishbone bursts. 
Measured signals are selected within a time window of 0.5 ms before and after the \textit{i}$^\textit{th}$ fishbone burst which is defined as the time point at which $|\Btheta|_{i}$ reaches its maximum value, i.e. $t_{i,\mathrm{ref}}=t_{i}(|\Btheta|_{i}=|\Btheta|_{i,\mathrm{max}})$, during the fishbone's evolution. The length of the time window is heuristically chosen to include sufficient data while minimizing the influence of background fluctuations. 
Because the fishbone activity increases before and decreases after $t_{i,\mathrm{ref}}$, its amplitude may become comparable to background fluctuations when the time window is too long.
This analysis supports determining which fluctuation, i.e., core or edge, is leading the other in terms of the phase. Although previous research~\cite{Wonjun_2023} demonstrated that both core and edge electron temperature fluctuations are coherent with magnetic fluctuations during the frequency-saturated phase, indicating that they are also coherent with each other, relative phases during fishbone bursts have not been addressed in that study. 

Measured signals are filtered in the frequency range of the fishbones, and their instantaneous phases are obtained using the Hilbert transform. With these instantaneous phases, a phase difference between the core and the edge fluctuations of electron temperature during a fishbone burst is calculated. If the fishbone activities are too weak, fishbone-relevant signals can be contaminated by other background fluctuations of the plasmas. We note that this can result in inaccurate instantaneous phase estimation. 

In order to substantiate if an observed phase difference is consistent across multiple fishbones, Lissajous curves between the core and the edge fluctuations of electron temperature are used. Parametric equations for Lissajous curves have a form of $L=(\frac{x(t)}{A},\frac{y(t)}{B})$ where $x(t)=A\sin(\omega_{1}t+\Delta \phi), y(t)=B\sin(\omega_{2}t)$. Here, $\Delta \phi$ is the phase of $x(t)$ with respect to $y(t)$. The shape of a Lissajous curve provides information on the frequency ratio, i.e., $\omega_{1}/\omega_{2}$, and the phase difference, i.e., $\Delta \phi$. In practice, $A$ and $B$ are time-varying, thus the signals are divided by their amplitude envelopes. Since the frequency of a fishbone chirps down, the frequency is also time-varying. For our case, we find based on the power spectra that $\omega_{1}$ and $\omega_{2}$ are identical to the frequency of a fishbone $\omega_{\mathrm{FB}}(t)$, i.e. $\omega_{1} = \omega_{2} = \omega_{\mathrm{FB}}(t)$. A reason why we are using Lissajous curves, instead of a cross-phase analysis, is that the frequency is chirping down when the fishbone bursts. 

For identical frequencies, $\omega_{1}=\omega_{2}$, a Lissajous curve exhibits the same shape regardless of whether the phase difference $\Delta \phi$ is positive or negative. The key difference between the two lies in the rotation direction of the Lissajous curve. Since it is inefficient to verify rotation directions from clouds of Lissajous curves, we additionally examine the distribution of $\sin\Delta \phi$. A positive (negative) value of $\sin\Delta \phi$ indicates that the phase difference is positive (negative). To further rectify the ambiguity of $\sin\Delta \phi$ due to its odd and periodic nature, we cross-check $\cos\Delta \phi$. For instance, if $\Delta \phi \sim 0$ or $\pm\pi$, then $\cos\Delta \phi \sim +1$ or $-1$ while $\sin\Delta \phi \sim 0$ always.

\subsection{Results}
\label{sec:results}

\subsubsection{Investigated KSTAR discharges}\label{sec:investigated_shots}
\begin{figure}
    [tb]\centering
    \includegraphics[width=0.45\textwidth,keepaspectratio]{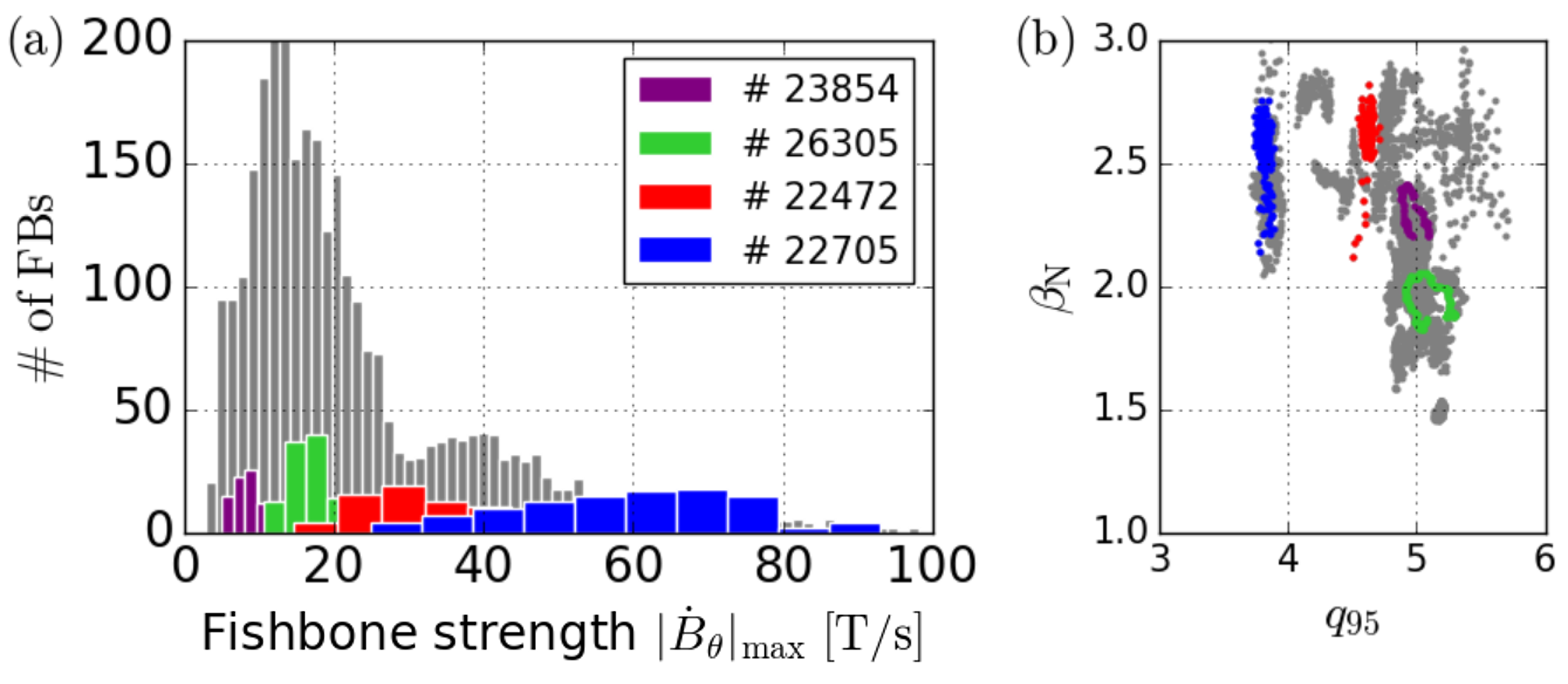}
    \caption{(a) Histograms of the double-peaked fishbones categorized into four groups depending on their strengths, and (b) their corresponding $\betaN$ and $\qedge$ values. Gray color indicates all the fishbones investigated in this work. Purple, green, red and blue indicate very weak, weak, moderate-strength and strong fishbones, respectively.}
    \label{fig:combined_stat}
\end{figure}

The double-peaked fishbones are categorized into four groups based on their strengths, and their representative KSTAR shots are shown in Figure \ref{fig:combined_stat}, together with their corresponding $\betaN$ and $\qedge$ values.

Very weak double-peaked fishbones (purple, $|\Btheta|_{\mathrm{max}} \lesssim 10$~T/s) are observed in the H-mode plasma with magnetic perturbations that suppress the ELM-crash with a density pump-out and energy confinement degradation in KSTAR shot \#23854. We have analyzed 79 of such weak fishbone bursts.

Weak fishbones (green, $10\lesssim |\Btheta|_{\mathrm{max}}\lesssim 20$~T/s) are observed in KSTAR shot \#26305. In KSTAR shot \#26305, the magnetic perturbations, causing a density pump-out and energy confinement degradation, are applied to suppress ELM-crashes. The total number of analyzed fishbones in KSTAR shot \#26305 is 113. Fishbones appear in the normalized beta $\betaN\sim 2$ and the edge safety factor $\qedge \sim 5$, as shown in Figure \ref{fig:combined_stat}(b). 

Moderate-strength fishbones (red, $20\lesssim |\Btheta| \lesssim 40$~T/s) are observed in KSTAR shot \#22472. The total number of analyzed fishbones in KSTAR shot \#22472 is 65. Most of the fishbones appear in the normalized beta $2.5\leq\betaN\leq2.8$ and the edge safety factor $\qedge \sim 4.6$, as shown in Figure \ref{fig:combined_stat}(b). In KSTAR shot \#22472, fishbones appear in the inter-ELM period. No external magnetic perturbation is applied in shot \#22472.

Strong fishbones (blue, $|\Btheta|_{\mathrm{max}}\gtrsim 40$~T/s) are observed in KSTAR shot \#22705. The total number of analyzed fishbones in KSTAR shot \#22705 is 105. Fishbones appear in the normalized beta $2.2\lesssim \betaN\lesssim 2.8$ and the edge safety factor $\qedge \sim 3.8$, as shown in Figure \ref{fig:combined_stat}(b). In KSTAR shot \#22705, the energy confinement is enhanced by 3D effects of the magnetic perturbations~\cite{Kimin_2023}. 

\begin{figure}
    [tb]\centering
    \includegraphics[width=0.35\textwidth,keepaspectratio]{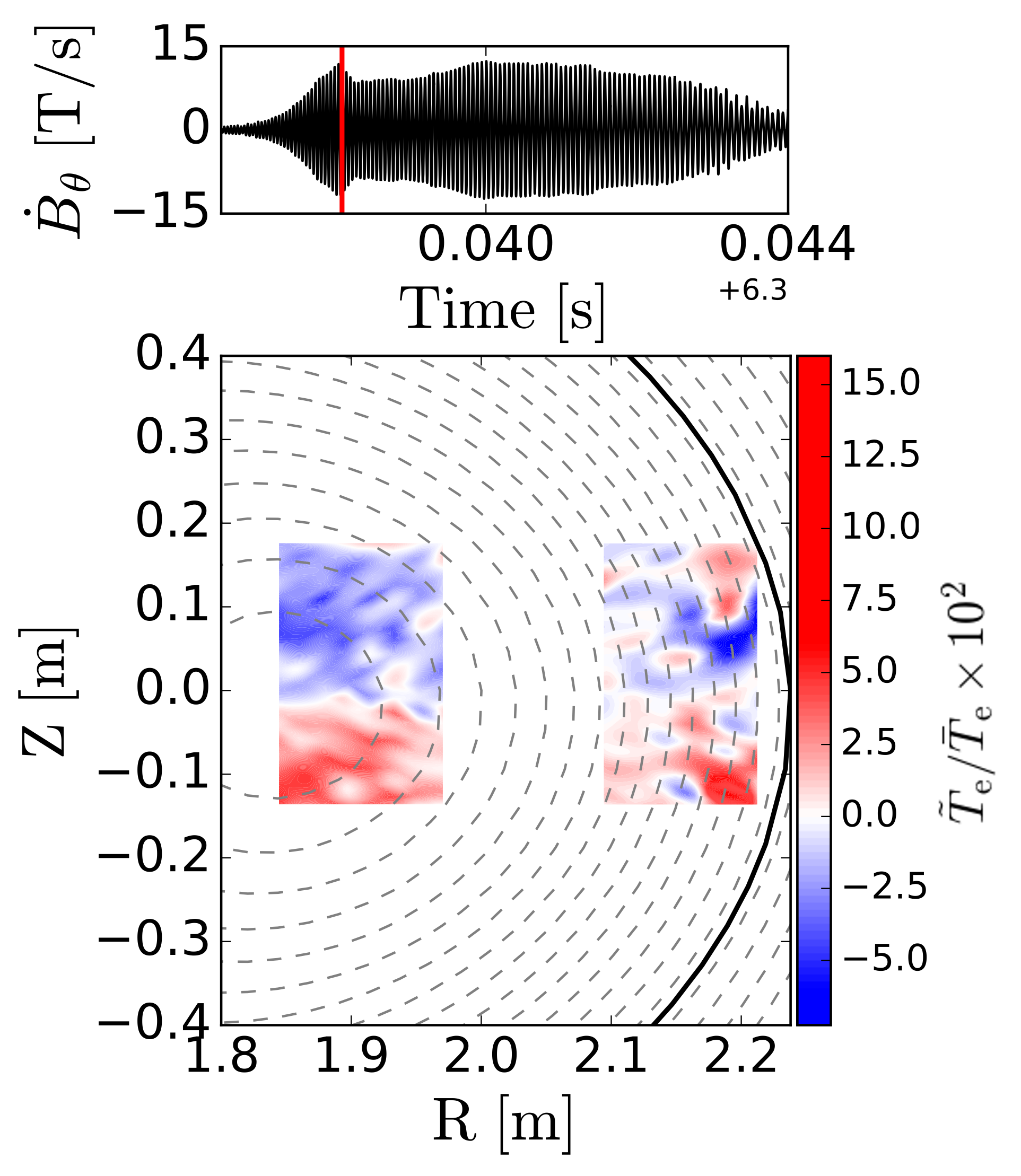}
    \caption{ECE image of a double-peaked fishbone observed in KSTAR shot \#26305 (weak fishbones). The upper panel shows the magnetic perturbations, with the red bar indicating the time at which the ECE image is displayed in the lower panel where the dashed lines represent the flux surfaces, and the black thick solid line is the last closed flux surface.}
    \label{fig:26305_ECEI}\vspace{-0.2cm}
\end{figure}

\begin{figure}
    [tb]\centering
    \includegraphics[width=0.35\textwidth,keepaspectratio]{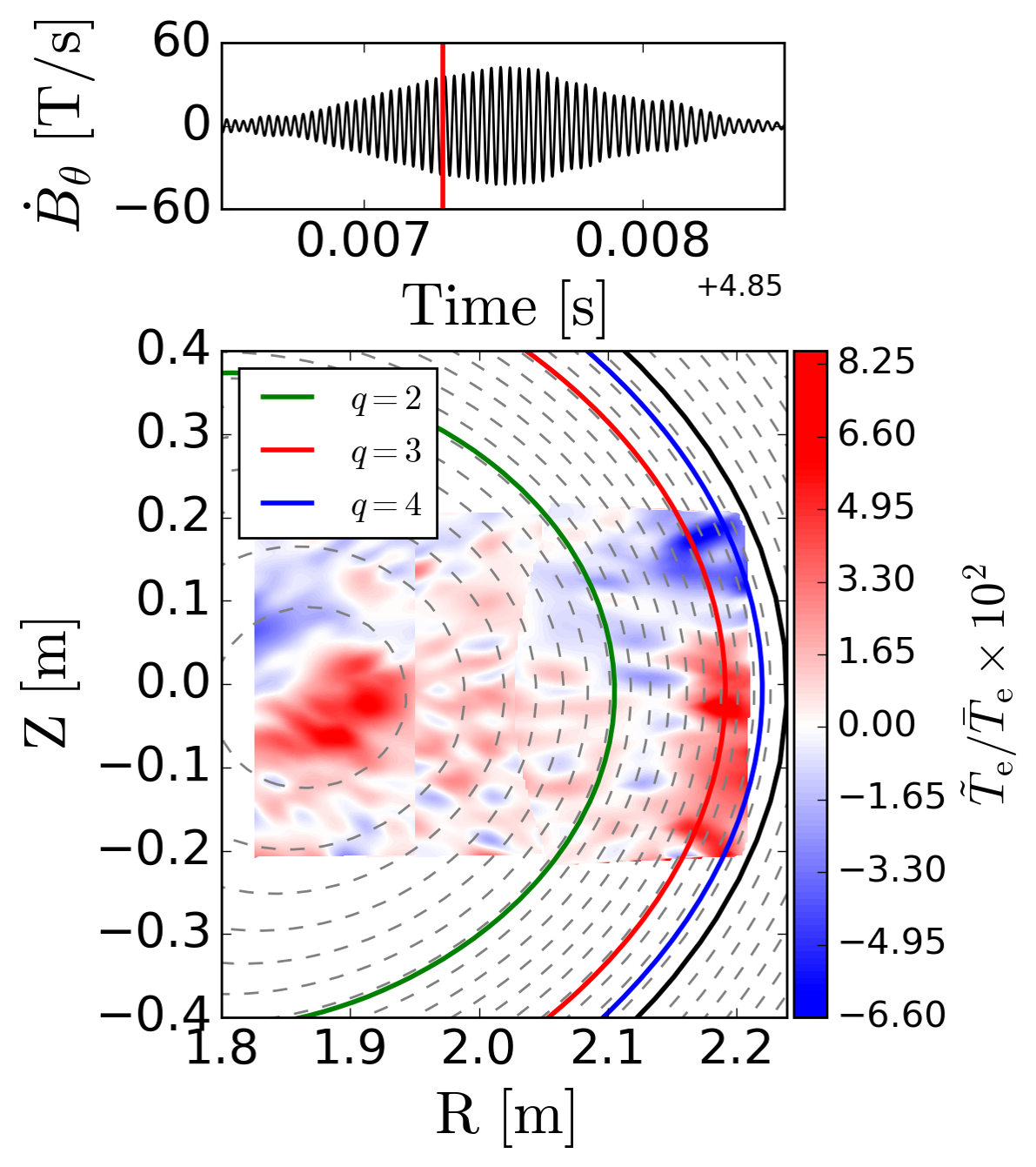}
    \caption{Same as Figure \ref{fig:26305_ECEI} in KSTAR shot \#22472 (moderate-strength fishbones).}
    \label{fig:22472_ECEI}
\end{figure}

\begin{figure}
    [tb]\centering
    \includegraphics[width=0.35\textwidth,keepaspectratio]{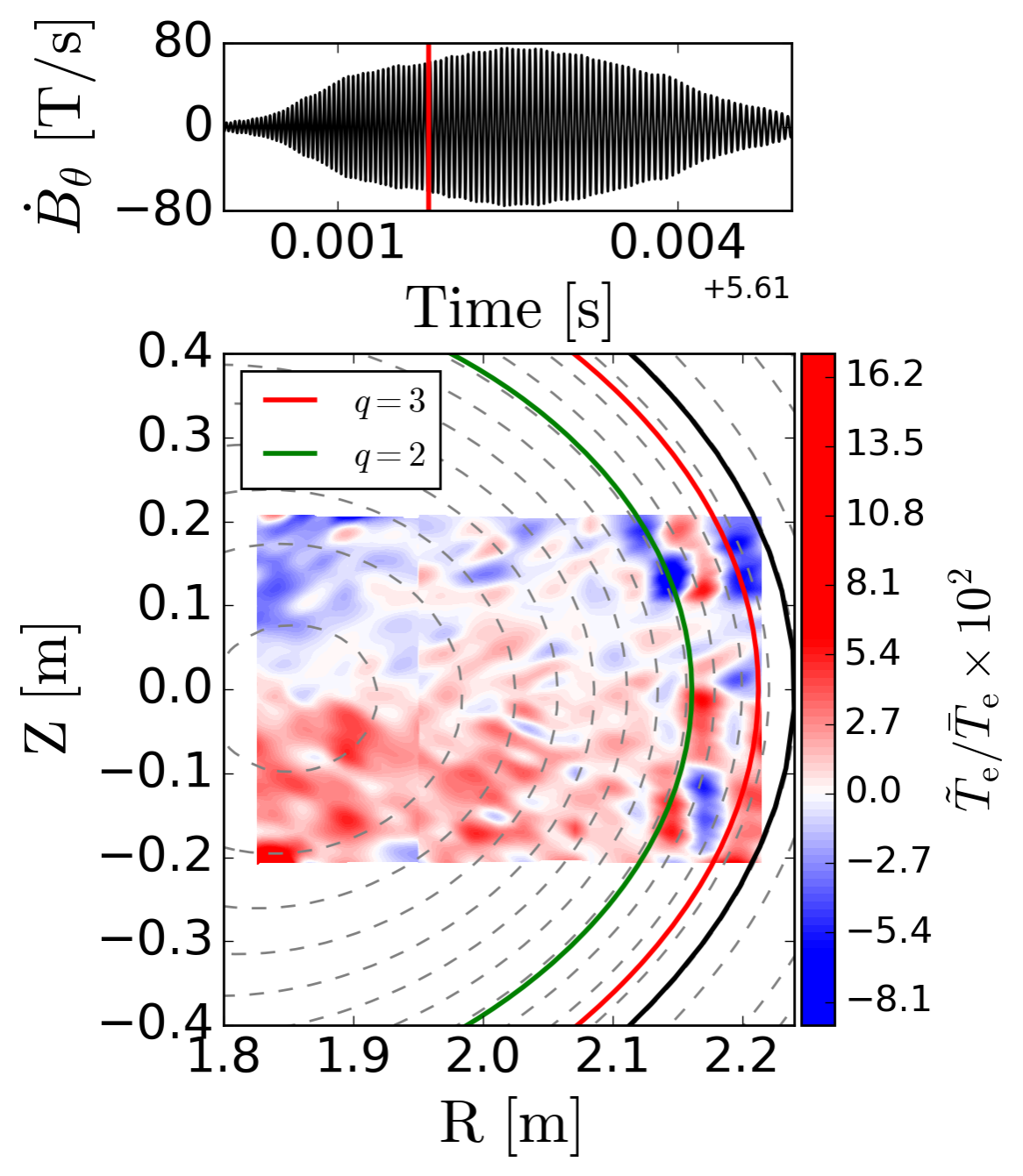}
    \caption{Same as Figure \ref{fig:26305_ECEI} in KSTAR shot \#22705 (strong fishbones).}
    \label{fig:22705_ECEI_pi_diff}\vspace{-0.2cm}
\end{figure}

The ECE images, corresponding to the relative temperature fluctuations $\Te/\bar{T_{\mathrm{e}}}$ where $\bar{T_{\mathrm{e}}}$ is an average electron temperature, of fishbone events are available in weak (shot \#26305, Figure \ref{fig:26305_ECEI}), moderate-strength (shot \#22472, Figure \ref{fig:22472_ECEI}) and strong fishbones (shot \#22705, Figure \ref{fig:22705_ECEI_pi_diff}). In Figures \ref{fig:26305_ECEI}-\ref{fig:22705_ECEI_pi_diff}, the top row shows the temporal evolution of the fluctuating poloidal magnetic fields associated with a fishbone event, where the red vertical bar indicates the time at which the ECE image, shown in the bottom row, is displayed. All presented ECE image data are band-pass filtered within the frequency range of fishbone activities. It is evident, for instance in Figure \ref{fig:26305_ECEI} representing a weak fishbone event, that both core and edge activities of a double-peaked fishbone are present. There is no ECEI data in the intermediate region ($2\lesssim R \lesssim 2.1$~m) in Figure \ref{fig:26305_ECEI}. Note that ECE image channels located from core to intermediate region, i.e. $1.8<R<2.1$~m, and those channels located at the edge region, i.e., $R>2.1$~m, are toroidally separated by $18.5^\circ$~\cite{KSTAR_ECEI_2014} for shots \#22472 (Figure \ref{fig:22472_ECEI}) and \#22705 (Figure \ref{fig:22705_ECEI_pi_diff}), while all the ECE image channels for shot \#26305 (Figure \ref{fig:26305_ECEI}) are located in the same toroidal position.

In shot \#22472, a clear ECE image is obtained during a moderately strong double-peaked fishbone event, as shown in Figure \ref{fig:22472_ECEI}. The green, red and the blue curves overlaid over the ECE image indicate $q=2$, $3$ and $4$ flux surfaces, respectively. Strong core and edge activities of a double-peaked fishbone are observed at $R\sim1.85$~m, and at the $q=3$ surface, respectively. 

The ECE image of a fishbone from shot \#22705, representing a strong fishbone event, is shown in Figure \ref{fig:22705_ECEI_pi_diff}. Here, the green and red curves in the ECE image are the flux surfaces at $q=2$ and $q=3$, respectively. A notable feature in this case is that there seems to exist a structure with radially alternating phase within the edge activity, i.e., the edge activities between the $q=2$ and $q=3$ surfaces. This feature is further addressed in Appendix \ref{appendix:pi_phase_diff}. Here, we argue that this should not be considered as a triple-peaked fishbone, but rather as a double-peaked fishbone consisting of core and edge activities. Such edge activities with a radially alternating phase structure is only observed sporadically in plasmas with strong fishbones, and similar features are sometimes interpreted as evidence for tearing-parity perturbations~\cite{Du_2021}.

\begin{figure}
    [tb]\centering
    \includegraphics[width=0.45\textwidth,keepaspectratio]{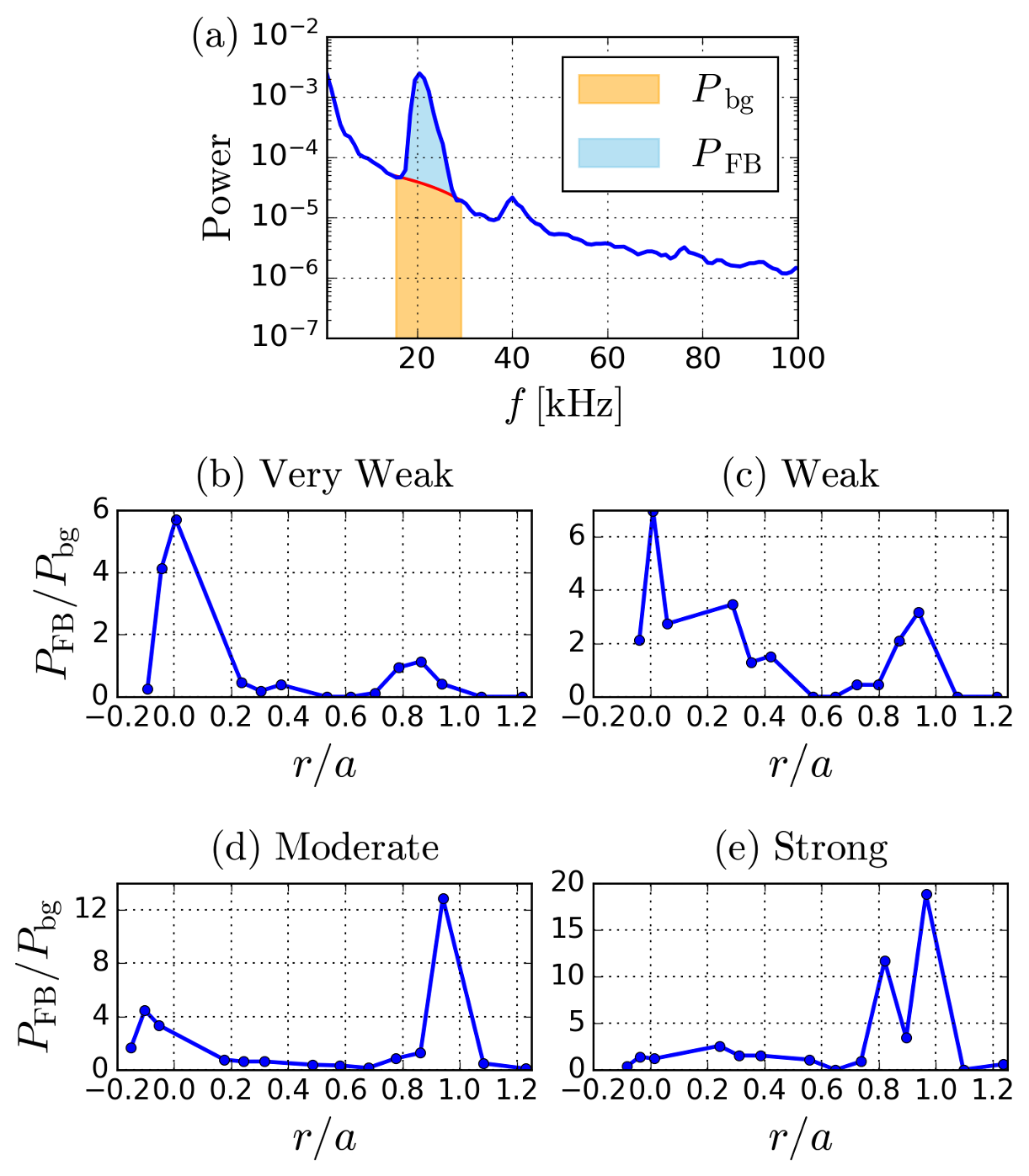}
    \caption{Power ratio between fishbone activities ($P_{\mathrm{FB}}$) and background fluctuations ($P_{\mathrm{bg}}$) for ECE channels of investigated shots. (a) An example of the power spectrum of the ECE signal (blue line), the interpolated power spectrum of background fluctuation (red line), $P_{\mathrm{FB}}$ (blue shaded area) and $P_{\mathrm{bg}}$ (orange shaded area).  The radial profile of power ratio $P_{\mathrm{FB}}/P_{\mathrm{bg}}$ : (b) very weak (\#23854), (c) weak (\#26305), (d) moderate-strength (\#22472), (e) strong (\#22705) fishbones.}
    \label{fig:FB_SNR}\vspace{-0.2cm}
\end{figure}

The power ratio between fishbone activities ($P_{\mathrm{FB}}$) and background fluctuations ($P_{\mathrm{bg}}$) for ECE radiometer channels in the investigated shots is illustrated in Figure \ref{fig:FB_SNR}. Figure \ref{fig:FB_SNR}(a) presents an example power spectrum (blue line) of the ECE signal averaged over many fishbone events as seen by one particular ECE channel. The peak at $20$~kHz corresponds to the fishbone activity. The background fluctuation level at the same frequency is interpolated and indicated by the red line. From this, the powers of fishbone activities ($P_{\mathrm{FB}}$, blue shaded area) and background fluctuations ($P_{\mathrm{bg}}$, orange shaded area) for every ECE channel are separately obtained. We note that this information should not be used to compare the power levels between different ECE channels such as comparing the absolute energy contents between fishbone's inner and outer components.

The radial profiles of the power ratio $P_{\mathrm{FB}}/P_{\mathrm{bg}}$ for (b) very weak (\#23854), (c) weak (\#26305), (d) moderate-strength (\#22472), (e) strong (\#22705) fishbones are shown in Figure \ref{fig:FB_SNR}. For very weak and weak fishbones, the ratio $P_{\mathrm{FB}}/P_{\mathrm{bg}}$ is approximately $6$ at the core and smaller at the edge. Specifically, for very weak fishbones, $P_{\mathrm{FB}}/P_{\mathrm{bg}}\sim 1$ at the edge, indicating that edge activity is comparable to background fluctuations. This limits the reliability of the phase analysis between core and edge activities for very weak fishbones (see Section \ref{sec:phase_analysis}). In contrast, for moderate-strength and strong fishbones, $P_{\mathrm{FB}}/P_{\mathrm{bg}}$ is significantly higher at the edge than at the core; specifically $P_{\mathrm{FB}}/P_{\mathrm{bg}}\gg 1$ for edge activity.

\subsubsection{Amplitude envelope components}

The amplitude envelope components of fishbone activities are analyzed through the conditional average and correlation coefficient, as described in section \ref{sec:anal_method}. For simplicity, we generally denote $\Te(r/a<0.4)$ by $\Te^{\Core}$, $\Te(r/a>0.7)$ by $\Te^{\Edge}$ and $\Te(0.4<r/a<0.7)$ by $\Te^{\Inter}$, while exact radial positions slightly vary for different KSTAR discharges. 

\begin{figure*}
    [!tb]\centering\vspace{-0.6cm}
    \makebox[\textwidth][c]{\includegraphics[width=1.0\textwidth,keepaspectratio]{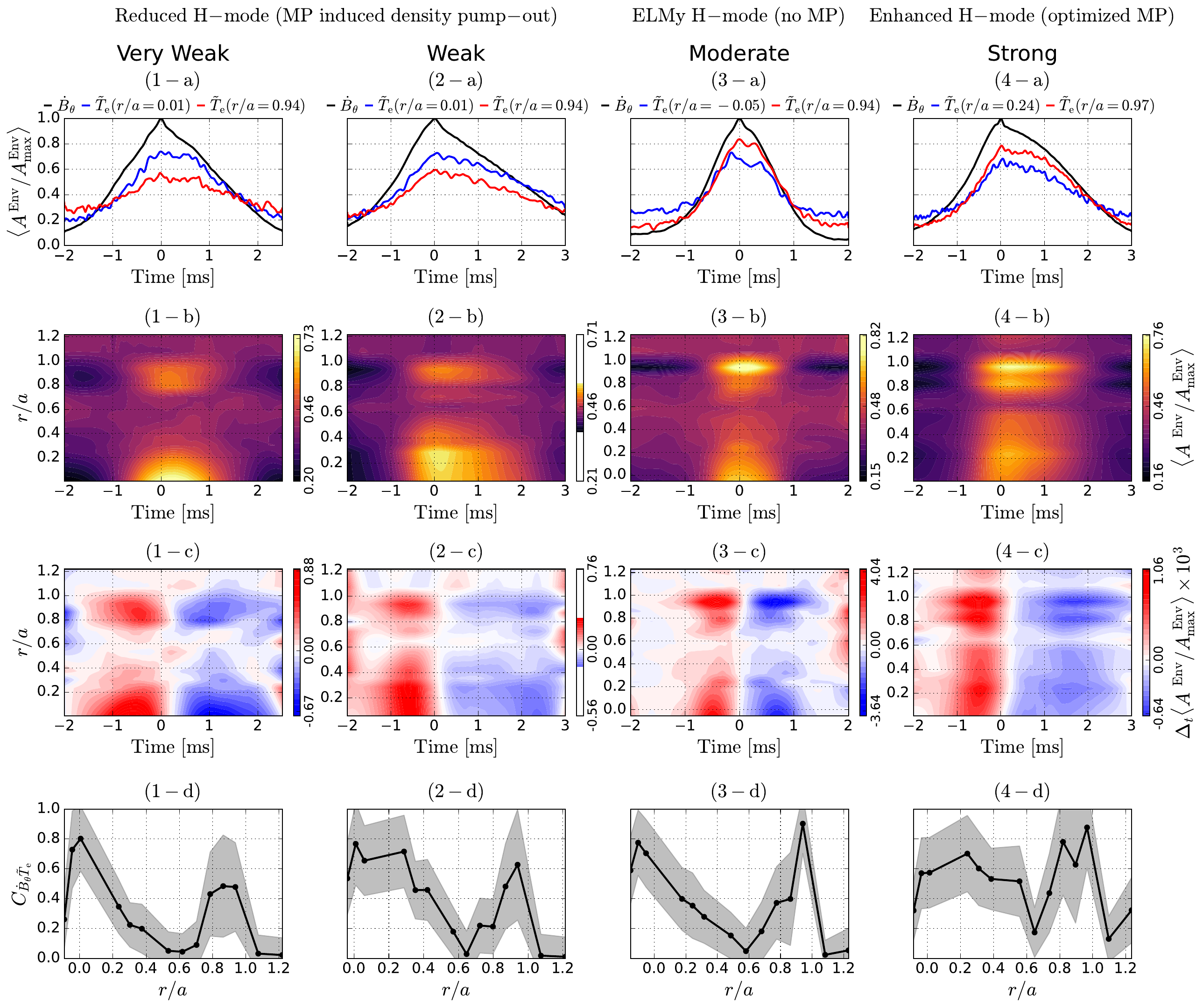}}
    \caption{Analyses results on the amplitude envelope components associated with the very weak (shot \#23854, first column), weak (shot \#26305, second column), moderate-strength (shot \#22472, third column) and strong (shot \#22705, fourth column)  fishbones. (1-a)-(4-a) Conditional average of normalized amplitude envelopes $\ConAvgNormAmpEnv$ for $\Btheta$ (black), $\Te^{\Core}$ (blue) and $\Te^{\Edge}$ (red) as a function of time. (1-b)-(4-b) Spatio-temporal evolutions of $\ConAvgNormAmpEnv$ for $\Te$ as a function of $r/a$ (ordinate) and time (abscissa), and their differentials $\Delta_{t}\ConAvgNormAmpEnv$ in (1-c)-(4-c). (1-d)-(4-d) Correlation coefficients between the normalized amplitude envelopes $C_{\Btheta \Te}$ of $\Btheta$ and $\Te$ as a function of $r/a$ with their standard deviations indicated by shades.}
    \label{fig:all_condavg_corr}
\end{figure*}

Figure \ref{fig:all_condavg_corr} shows all the results we have obtained from the amplitude envelope components for the very weak (shot \#23854, first column), weak (shot \#26305, second column), moderate-strength (shot \#22472, third column) and strong (shot \#22705, fourth column) fishbones. 

\begin{figure}
    [tb]\centering
    \includegraphics[width=0.45\textwidth,keepaspectratio]{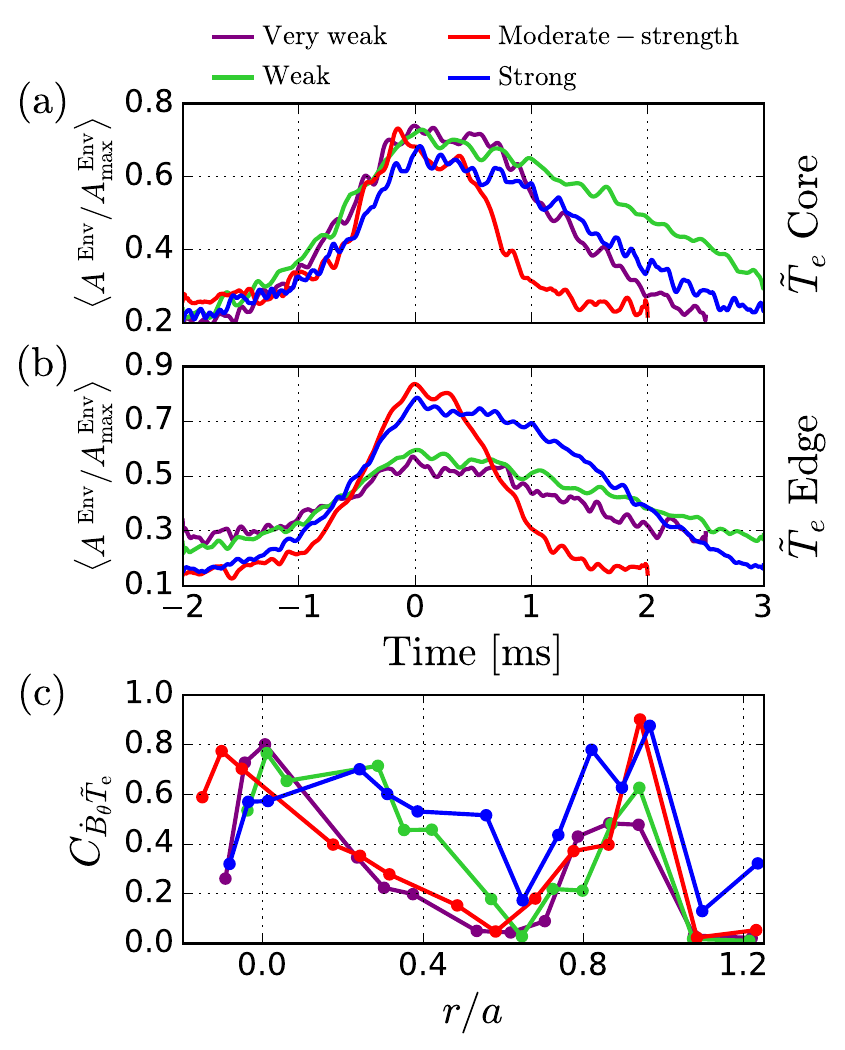}
    \caption{Results shown in Figure \ref{fig:all_condavg_corr} are rearranged for easier comparisons among the very weak (shot \#23854, purple), weak (shot \#26305, green), moderate-strength (shot \#22472, red) and strong (shot \#22705, blue) fishbones. The conditional average of normalized amplitude envelopes $\ConAvgNormAmpEnv$ for (a) $\Te^{\Core}$ and (b) $\Te^{\Edge}$, and (c) correlation coefficient $C_{\Btheta \Te}$.}
    \label{fig:summary_condavg_corr}
\end{figure}

Figure \ref{fig:all_condavg_corr}(1-a)-(4-a) show the conditional average of normalized amplitude envelopes, denoted as $\ConAvgNormAmpEnv$,  for $\Btheta$ (black), $\Te^{\Core}$ (blue) and $\Te^{\Edge}$ (red) as a function of time. As mentioned earlier, $t=0$ is defined as the time at which the amplitude envelope ($A^\textrm{Env}$) of $\Btheta$ reaches its maximum. The maxima of $\ConAvgNormAmpEnv$ for the core activities stay relatively constant as the fishbone activities become stronger while those for the edge activities become not only larger compared to itself (see Figure \ref{fig:summary_condavg_corr}(a) and (b)) but also greater than the core activities. This means that it is the edge activities that more affect or are more affected by the strength of the fishbones compared to the core activities in terms of the conditional average of normalized amplitude envelopes $\ConAvgNormAmpEnv$.

For completeness, we also present spatio-temporal evolutions of $\ConAvgNormAmpEnv$ for $\Te$, i.e., contour as a function of $r/a$ (ordinate) and time (abscissa), using all the radial channels of the ECE radiometer in Figure \ref{fig:all_condavg_corr}(1-b)-(4-b). We clearly observe two peaks at the core ($r/a \lesssim 0.4$) and edge ($r/a\sim 0.8$) region, while there are minimal fishbone activities in the intermediate region ($0.4<r/a<0.7$) for very weak, weak and moderate-strength fishbones. In case of the strong fishbones, the core and edge components of the mode are separated by a narrow but distinct minimum around $r/a=0.6$. In addition, in this particular example of strong fishbone activities shown in Figure \ref{fig:all_condavg_corr}(4-b), there are two distinct sub-peaks within the edge region, which is due to the radially alternating phase structure as seen in Figure \ref{fig:22705_ECEI_pi_diff}. Again, this is further elaborated in Appendix \ref{appendix:pi_phase_diff}.

Similarly, contours of $\Delta_{t}\ConAvgNormAmpEnv$ are also shown in Figure \ref{fig:all_condavg_corr}(1-c)-(4-c). For all four cases, the core and the edge activities start to grow and decay almost simultaneously. Although statistically not yet verified, we find that the edge activity persists slightly longer than the core one only in the case of the strong fishbone as seen in Figure \ref{fig:all_condavg_corr}(4-b) and (4-c). Statistical analyses on the durations of the fishbones will be investigated as future work.

Correlation coefficients between the normalized amplitude envelopes of $\Btheta$ and $\Te$, denoted as $C_{\Btheta \Te}$, along the radial ECE radiometer channels are shown in Figure \ref{fig:all_condavg_corr}(1-d)-(4-d). They are estimated using Eq. (\ref{eq:corr}), and Eq. (\ref{eq:corr_sd}) is used to calculate the standard deviation. They clearly show the double peaks, which is consistent with the coherence analyses and the fluctuation level comparisons reported in a previous work \cite{Wonjun_2023}. Similar to $\ConAvgNormAmpEnv$, i.e., Figure \ref{fig:all_condavg_corr}(1-a)-(4-a), core correlation levels do not change much for four different strengths of the fishbone events, while the edge correlation levels increase with the strengths (see Figure \ref{fig:summary_condavg_corr}(c)).

Results shown in Figure \ref{fig:all_condavg_corr} are rearranged in Figure \ref{fig:summary_condavg_corr} for easier comparisons among different strengths of the fishbone activities. It is clear in Figure \ref{fig:summary_condavg_corr}(a) that the peak values of $\ConAvgNormAmpEnv$ for $\Te^{\Core}$ do not depend on the strengths of the fishbone activities, while those for $\Te^{\Edge}$ monotonically increase with the strength as shown in Figure \ref{fig:summary_condavg_corr}(b). Likewise, Figure \ref{fig:summary_condavg_corr}(c) shows similar trends for $C_{\Btheta \Te}$.

\subsubsection{Phase components}\label{sec:phase_analysis}

\begin{figure*}
    [!tb]\centering\vspace{-0.5cm}
    \makebox[\textwidth][c]{\includegraphics[width=1.0\textwidth,keepaspectratio]{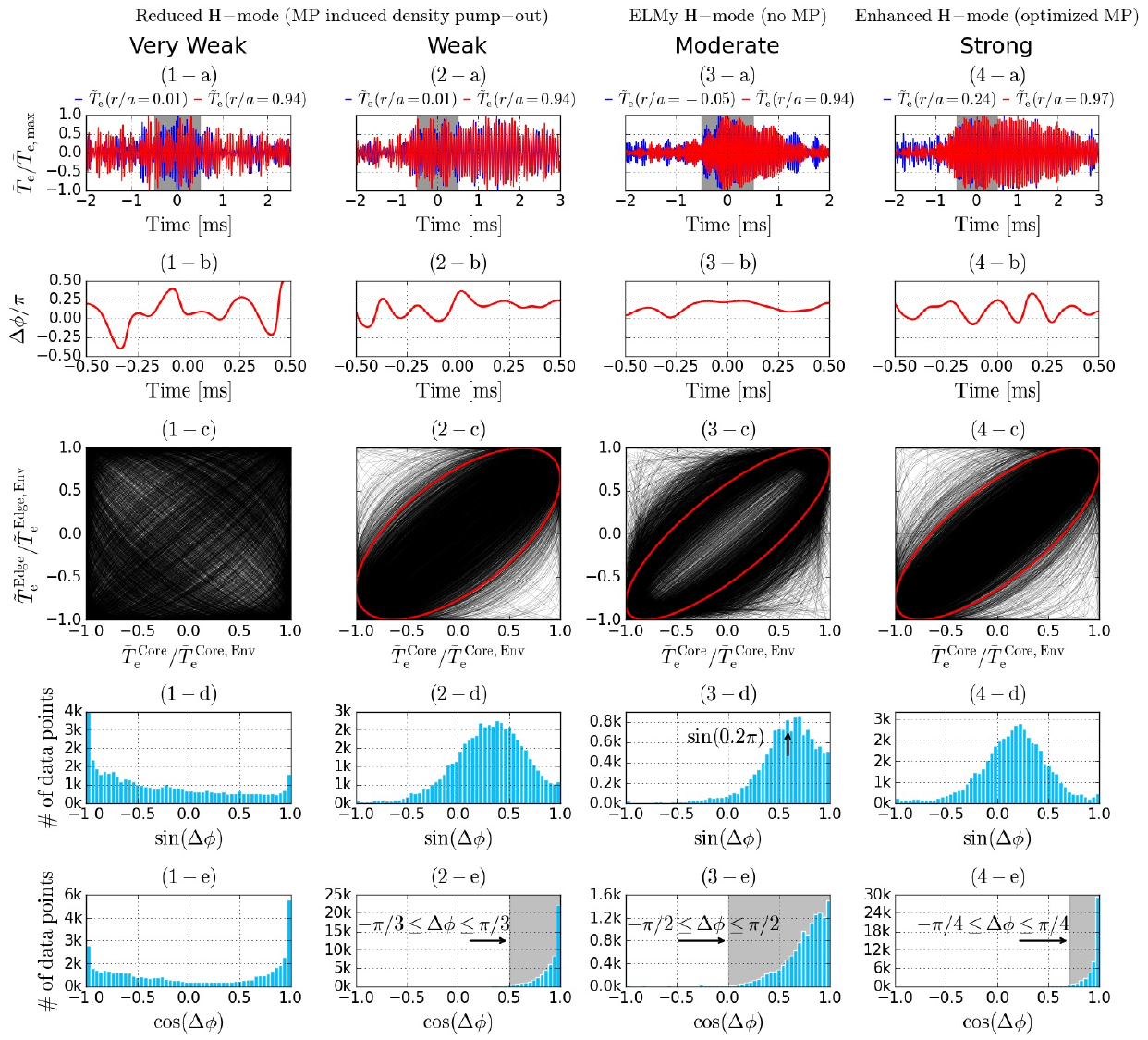}}
    \caption{Analyses results on the phase components associated with the very weak (shot \#23854, first column), weak (shot \#26305, second column), moderate-strength (shot \#22472, third column) and strong (shot \#22705, fourth column) fishbones. (1-a)-(4-a) Exemplary temporal evolutions of normalized $\Te^\Core$ (red) and $\Te^\Edge$ (blue) of a single fishbone event measured by the ECE radiometer. (1-b)-(4-b) Instantaneous phase difference $\Delta \phi = \phi(\Te^{\Edge})-\phi(\Te^{\Core})$ within the temporal range indicated by the gray shaded areas in (1-a)-(4-a), i.e., $t=[-0.5, 0.5]$~ms. Note that $t=0$ is defined the same as in the case of the amplitude envelope analyses shown in Figure \ref{fig:all_condavg_corr}. (1-c)-(4-c) Lissjous curves $L=\left (\frac{\Te^{\Core}}{\Te^{\Core,\Env}},\frac{\Te^{\Edge}}{\Te^{\Edge, \Env}} \right )$ of ensembles of the fishbone events for $t=[-0.5, 0.5]$~ms with the red Lissajous curves corresponding to $\Delta\phi=0.28\pi$ in (2-c), $0.2\pi$ in (3-c) and $0.23\pi$ in (4-c) whose meanings are described in texts. With the same datasets used for Lissajous curves, histograms of $\sin(\Delta\phi)$ and $\cos(\Delta\phi)$ are shown in (1-d)-(4-d) and (1-e)-(4-e), respectively, demonstrating that $\Delta\phi$ are more likely (in a statistical sense) to persist in the range of $0\leq \Delta\phi \leq \pi/3$ for weak, $0 \leq \Delta\phi \leq \pi/2$ for moderate-strength, and $0 \leq \Delta\phi \leq \pi/4$ for strong fishbone activities.}
    \label{fig:all_phase_analysis}
\end{figure*}

We have also scrutinized phase differences between fluctuating temperatures associated with the fishbone events from the core ($\Te^{\Core}$) and edge ($\Te^{\Edge}$) measured by the ECE radiometer. Similar to Figure \ref{fig:all_condavg_corr}, all the results obtained by examining the phases are presented in Figure \ref{fig:all_phase_analysis} for four different groups, i.e., very weak (shot \#23854, first column), weak (shot \#26305, second column), moderate-strength (shot \#22472, third column) and strong (shot \#22705, fourth column) fishbones. 

As examples, Figure \ref{fig:all_phase_analysis}(1-a)-(4-a) show temporal evolutions of normalized $\Te^{\Core}$ (blue) and $\Te^{\Edge}$ (red) of a single double-peaked fishbone event from four different groups. Here, the signals are normalized by their own maximum amplitudes. 

An instantaneous phase $\phi$ of the individual signal within the temporal range indicated by the gray shaded areas in Figure \ref{fig:all_phase_analysis}(1-a)-(4-a), i.e., $-0.5\leq t \leq 0.5$~ms where $t=0$ is defined the same as in the case of the amplitude envelope analyses, is obtained using the Hilbert transform. Then, the phase difference $\Delta\phi$ between $\Te^{\Core}$ and $\Te^{\Edge}$ is calculated as $\Delta \phi = \phi(\Te^{\Edge})-\phi(\Te^{\Core})$, which is shown in Figure \ref{fig:all_phase_analysis}(1-b)-(4-b). In the case of the very weak fishbones (first column), the phase difference fluctuates between $-0.5\pi$ and $0.5\pi$; whereas for weak, moderate-strength and strong fishbones the phase differences stay above zero and fluctuate between $0$ and $0.3\pi$. This suggests that, in these examples, the phase of $\Te^{\Edge}$ precedes that of $\Te^{\Core}$ during the fishbone events. 

For a statistical approach, we extend the phase analysis to ensembles of fishbones within the time range of $-0.5\leq t \leq 0.5$~ms for each fishbone and draw the Lissajous curves. The Lissajous curves $L=\left (\frac{\Te^{\Core}}{\Te^{\Core,\Env}},\frac{\Te^{\Edge}}{\Te^{\Edge, \Env}} \right )$ associated with the fishbone events are shown in Figure \ref{fig:all_phase_analysis}(1-c)-(4-c). Here, $\Te^{\Core,\Env}$ and $\Te^{\Edge,\Env}$ are amplitude envelopes of $\Te^\Core$ and $\Te^\Edge$, respectively, which isolate the amplitude information out from the Lissajous curves, allowing us to examine only the phase information. We find, except in the very weak fishbone case, that the Lissajous curves in Figure \ref{fig:all_phase_analysis}(2-c)-(4-c), where instantaneous $\Delta \phi$ stays above zero during a fishbone event, form elliptical disks or clouds of ellipses, indicating a consistent phase difference across multiple fishbone events. For the very weak fishbone activities as in Figure \ref{fig:all_phase_analysis}(1-c), the Lissajous curves are covering all the range $-1\leq \frac{\Te^{\Core}}{\Te^{\Core,\Env}} \leq 1$ and $-1 \leq \frac{\Te^{\Edge}}{\Te^{\Edge,\Env}} \leq 1$, demonstrating the absence of consistent phase relations between $\Te^{\Core}$ and $\Te^{\Edge}$ during multiple fishbone events.

In addition, the sines and cosines of the phase differences $\Delta \phi = \phi(\Te^{\Edge})-\phi(\Te^{\Core})$ are calculated with the same datasets used for Lissajous curves, and their distributions are shown in Figure \ref{fig:all_phase_analysis}(1-d)-(4-d) and (1-e)-(4-e), respectively. Similar to the Lissajous curves, we observe no consistent $\Delta\phi$ for very weak fishbone events as shown in Figure \ref{fig:all_phase_analysis}(1-d) and (1-e), while there exist statistically significant features for weak, moderate-strength and strong fishbone events. It is worth noting that, for very weak fishbones, the edge activity is comparable to background fluctuations (see Section \ref{sec:investigated_shots}), thereby complicating accurate phase analysis. 

The gray-shaded regions in the distributions of $\cos(\Delta\phi)$, shown in Figure \ref{fig:all_phase_analysis}(2-e)-(4-e), indicate that the phase differences are concentrated around zero, i.e. $-\pi/3 \leq \Delta\phi \leq \pi/3$ for weak, $-\pi/2 \leq \Delta\phi \leq \pi/2$ for moderate-strength, and $-\pi/4 \leq \Delta\phi \leq \pi/4$ for strong fishbone events. We also find that $\sin(\Delta\phi)$, shown in Figure \ref{fig:all_phase_analysis}(2-d)-(4-d), are skewed towards the positive values. Combining observations from $\sin(\Delta\phi)$ and $\cos(\Delta\phi)$, the phase differences, $\Delta \phi = \phi(\Te^{\Edge})-\phi(\Te^{\Core})$, are more likely (in a statistical sense) to persist in the range of $0\leq \Delta\phi \leq \pi/3$ for weak, $0 \leq \Delta\phi \leq \pi/2$ for moderate-strength, and $0 \leq \Delta\phi \leq \pi/4$ for strong fishbone events. Again, $\Delta\phi > 0 $ means that the phase of $\Te^\Edge$ leads that of $\Te^\Core$.

We have drawn the red Lissajous curves on Figure \ref{fig:all_phase_analysis}(2-c)-(4-c) that represent the phase differences of $0.28\pi$, $0.2\pi$ and $0.23\pi$, respectively. Note that Lissajous curves with negative phase differences would be identical, but we only indicate the positive values since $\sin(\Delta\phi)$ is mostly positive (see Figure \ref{fig:all_phase_analysis}(2-d)-(4-d)). The red Lissajous curves in Figure \ref{fig:all_phase_analysis}(2-c) and (4-c) contain 90\% of black Lissajous curves, indicating that the phase differences, $\Delta \phi = \phi(\Te^{\Edge})-\phi(\Te^{\Core})$, are mostly limited within the range of $0\leq \Delta\phi \leq 0.28\pi$ and $0\leq \Delta\phi \leq 0.23\pi$ for weak and strong fishbone events, respectively. For moderate fishbones shown in Figure \ref{fig:all_phase_analysis}(3-c), black Lissajous curves form clouds of ellipses near the red Lissajous curves, i.e., $\Delta\phi=0.2\pi$, which is a location of the peak in Figure \ref{fig:all_phase_analysis}(3-d), corresponding to $\sin(0.2\pi)=0.7$.

\begin{figure}
    [tb]\centering
    \includegraphics[width=0.4\textwidth,keepaspectratio]{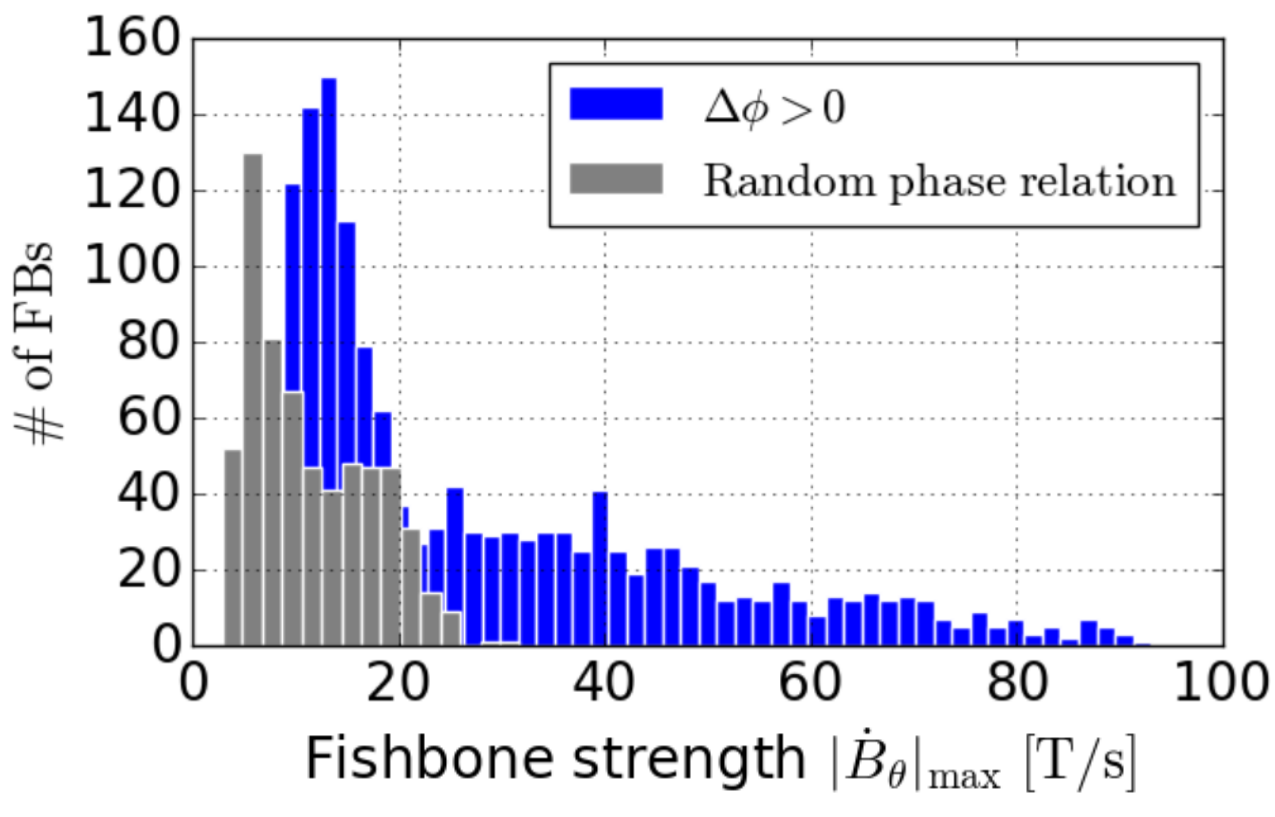}
    \caption{Distribution of fishbone strength is shown for two groups: one with a random phase difference between $\Te^{\Core}$ and $\Te^{\Edge}$ (gray) and the other with a positive phase difference, $\Delta \phi = \phi(\Te^{\Edge}) - \phi(\Te^{\Core})>0$ (blue). A total of 28 KSTAR discharges are analyzed.}
    \label{fig:trend_of_phase_relation}
\end{figure}

Such a trend of the phase relation, i.e., the phase of $\Te^\Edge$ leading that of $\Te^\Core$ except for very weak fishbone events, is also observed in many other KSTAR discharges. We have examined 28 KSTAR discharges, and Figure \ref{fig:trend_of_phase_relation} shows the distribution of the fishbone strength categorized into two groups: one with random phase difference (gray) and the other with positive phase difference (blue), i.e. $\Delta \phi = \phi(\Te^{\Edge}) - \phi(\Te^{\Core})>0$. Cases with the random phase difference are localized in the range of $|\Btheta|_{\mathrm{max}}\lesssim 20$~T/s, while those with definite positive phase difference are observed widely in the range of $|\Btheta|_{\mathrm{max}}\gtrsim 10$~T/s. 

At this point, we do not claim that very weak fishbone events have different phase relations between $\Te^\Core$ and $\Te^\Edge$ from stronger ones. This is because we cannot reject the possibility that the observation of the random $\Delta\phi$ could be just a consequence of very weak fishbone activities not protruding enough from other background fluctuations within a similar frequency range such as broadband turbulence. We emphasize that the observation that the phase of $\Te^{\Edge}$ leads that of $\Te^{\Core}$ for strong enough fishbone activities is statistically significant, i.e., not a random event. However, one must also keep in mind that the measurements were performed only near the outer (low-field side) midplane. The relative phases are expected to vary below and above the midplane if one makes the (reasonable) assumption that the inner and outer components of a double-peaked fishbone have different dominant poloidal mode numbers $m\sim nq$ (with $n=1$). This is discussed in more detail in Section 5.2.2 of the second paper in this series~\cite{Bierwage2026}.

\subsubsection{Fishbones in a plasma with a flat edge profile}\label{sec:notes_fishbones}

\begin{figure}
    [tb]\centering
    \includegraphics[width=0.4\textwidth,keepaspectratio]{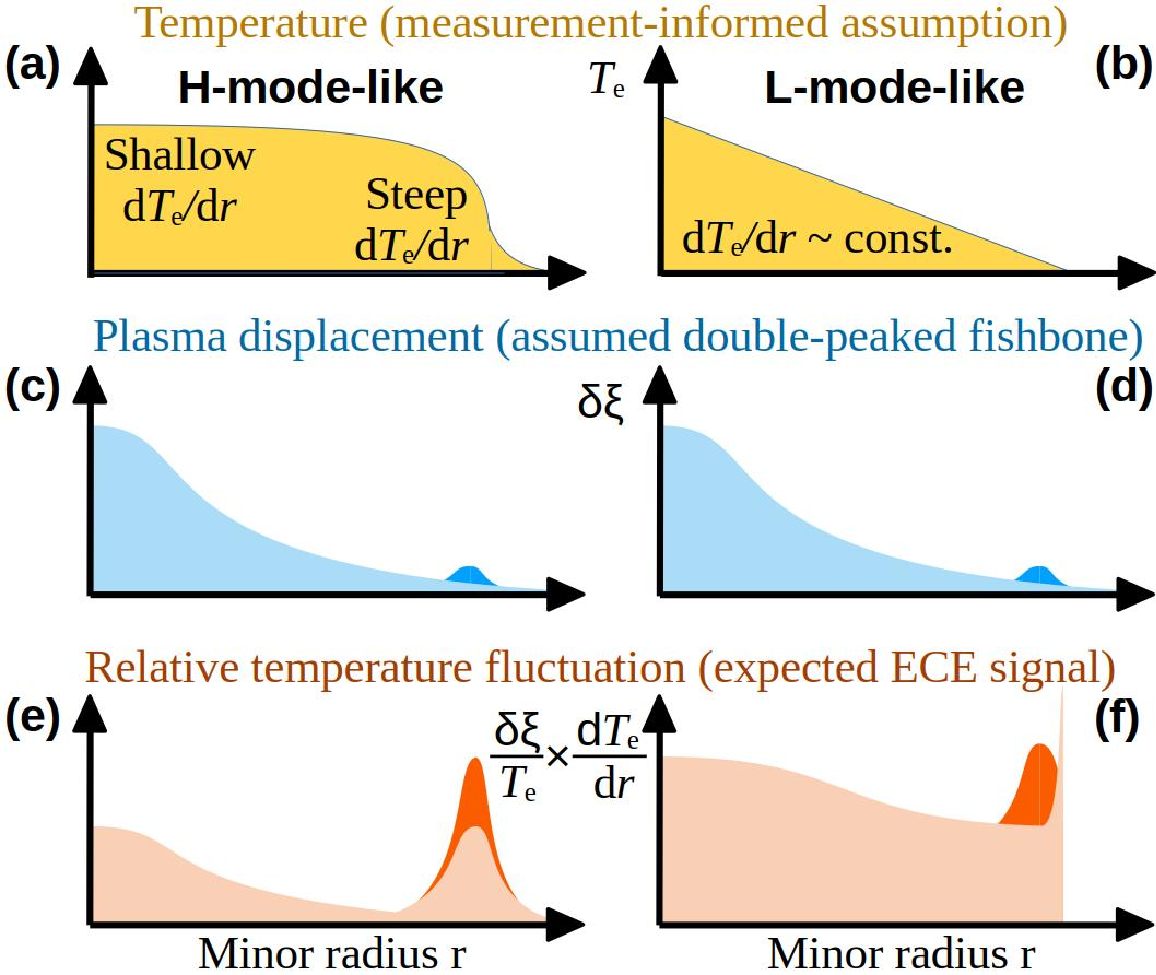}
    \caption{Schematic illustration of the influence of the temperature profile on the apparent form of a mode. Panels (a) and (b) show ``H-mode-like'' and ``L-mode-like'' temperature profiles $T_\mathrm{e}(r)$. Panels (c) and (d) show the radial displacement $\delta\xi$ caused by a hypothetical mode (light blue) that peaks near the axis and has a small but nonzero amplitude near the edge. The mode may have a true minor second peak near the edge (dark blue). For the two $T_\mathrm{e}$ profiles in (a) and (b), panels (e) and (f) show the expected apparent profile $\Te/T_\mathrm{e}$ of such a mode without (light red) and with true outer peak (dark red).}
    \label{fig:HL_mode_profs}
\end{figure}

Since all instances of double-peaked fishbones analyzed above and in Ref \cite{Wonjun_2023} were found in H-mode plasmas, and since all observations are based on relative density and/or temperature fluctuation in the form
\begin{equation}
\label{eq:relative_fluc}
\frac{\Te}{T_\mathrm{e}} \sim \delta\xi \frac{dT_\mathrm{e}/dr}{T_\mathrm{e}},
\end{equation}
where $\delta\xi$ is the radial displacement associated with the fluctuations, one may wonder whether the outer peak is merely a profile effect. This is illustrated schematically in the left column of Figure \ref{fig:HL_mode_profs}. In panel (a), we assume that $T_\mathrm{e}$ has a steep radial gradient $dT_\mathrm{e}/dr$ near the edge and a shallow gradient elsewhere. Regardless of whether the radial displacement $\delta\xi$ associated with the fishbone mode in Figure \ref{fig:HL_mode_profs}(c) has an edge peak (dark blue) or not (light blue), the profile of the measured relative temperature fluctuation $\Te/T_\mathrm{e}$ can always have a distinct peak near the edge due to the factor $(dT_\mathrm{e}/dr)/T_\mathrm{e}$ in Eq. (\ref{eq:relative_fluc}). This is illustrated in Figure \ref{fig:HL_mode_profs}(e). In contrast, if $T_\mathrm{e}$ has a more uniform gradient like Figure \ref{fig:HL_mode_profs}(b), an edge peak in the profile of $\Te/T_\mathrm{e}$ as in panel (f) is a strong indication that the radial displacement of the fishbone in panel (d) has a true edge peak as well.

\begin{figure}
    [tb]\centering
    \includegraphics[width=0.4\textwidth,keepaspectratio]{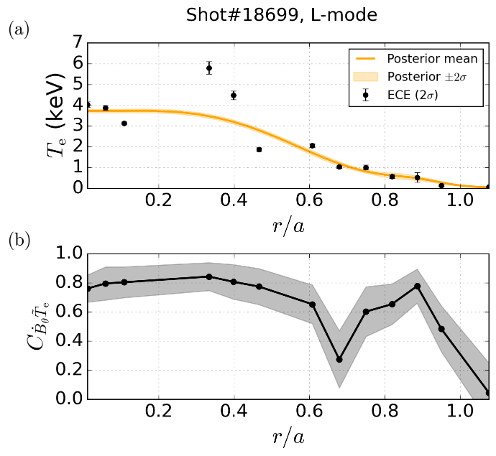}
    \caption{An example of the double-peaked fishbone events with a flat edge profile obtained from KSTAR shot \#18699 $t=1.3-1.7$~s. (a) The temperature profile from the ECE radiometer (circles) and the curve fit result (yellow) with the fit uncertainties (light yellow) obtained from many possible MCMC (Markov Chain Monte Carlo)-generated curve fits to the measured data points, and (b) the correlation coefficients between the normalized amplitude envelopes $C_{\Btheta \Te}$ of $\Btheta$ and $\Te$ as a function of $r/a$ with their standard deviations indicated by shades, clearly showing the outer component even with the L-mode-like edge profile.}
    \label{fig:FB_with_flat_profile}
\end{figure}

Thus motivated, we have scrutinized many KSTAR discharges that are not classified as H-modes and searched for instances of double-peaked fishbones there. An example is presented in Figure \ref{fig:FB_with_flat_profile}, where panel (a) shows the temperature profile measured by the ECE radiometer together with the curve fit, and panel (b) shows the correlation coefficients $C_{\Btheta \Te}$ as a function of $r/a$, obtained from a set of fishbone in KSTAR shot \#18699 $t=1.3-1.7$~s. We observe a broad inner peak and a somewhat narrower outer peak with high correlation coefficients. The two peaks are separated by a narrow but distinct region of reduced correlation around $r/a\sim 0.6-0.7$.

The situation resembles most closely that in Figure \ref{fig:all_condavg_corr}(4-d), where results of a similar analysis was shown in H-mode fishbone. However, this does not necessarily mean that Figure \ref{fig:FB_with_flat_profile}(b) shows an instance of what we classified as ``strong fishbones'' in Figure \ref{fig:all_condavg_corr}, because the form of the temperature profile can have an influence on the apparent width of the two peaks as illustrated in Figure \ref{fig:HL_mode_profs}(e) and (f). 

Among the surveyed discharges with L-mode-like edge profiles, the fishbones in KSTAR shot \#18699 analyzed in Figure \ref{fig:FB_with_flat_profile} have the largest outer mode component. This shall suffice here because we merely need one counter-example to rule out the necessity of the profile effect. It must be noted that fishbone activities in KSTAR plasmas with L-mode-like edge profiles as in Figure \ref{fig:FB_with_flat_profile}(a) are quite scarce. Nevertheless, Figure \ref{fig:FB_with_flat_profile} constitutes evidence in support of our assumption that we are dealing with truly double-peaked fishbones and that the profile effect in Eq. (\ref{eq:relative_fluc}) and Figure \ref{fig:HL_mode_profs} is unlikely to be a sufficient explanation for our observations, even though the $T_\mathrm{e}$ profile does have an influence on the apparent form of the modes.

\section{Discussion and Summary}\label{sec:summary}

As the double-peaked fishbones have been reported only from KSTAR~\cite{Wonjun_2023}, we have presented, in this work, detailed statistical characteristics of them based on 40 KSTAR discharges containing approximately 3,000 fishbone events in terms of observational conditions and spatio-temporal structures. Here, the double-peaked fishbone means a fishbone-like burst and chirp with a double-peaked mode structure in the radial direction, consisting of coherent core and edge activities with weaker or even no detectable activities in the intermediate region.

The maximum strength of the poloidal magnetic fluctuations associated with the double-peaked fishbones, denoted as $|\Btheta|_{\mathrm{max}}$ (biased to the strength of the edge activity), is observed to be highly correlated with the normalized beta $\betaN$ and the edge safety factor at the 95\% normalized flux surface $\qedge$, that is the strength tends to be larger for a larger $\betaN$ and a smaller $\qedge$. Such correlations are also associated with the presence and form of external magnetic perturbations since we observe the weakest (strongest) fishbone activities when externally applied magnetic perturbations degrade (enhance) the energy confinement in H-mode plasmas. Moderate-strength fishbone is observed in H-mode plasmas without any magnetic perturbations. 

As there exist correlations between the strength of fishbones and operation conditions, i.e., the presence and form of external magnetic perturbations, of KSTAR plasmas, statistical investigations on spatio-temporal structures are performed after classifying the double-peaked fishbones into four groups based on their strengths: very weak ($|\Btheta|_{\mathrm{max}} \lesssim 10$~T/s), weak ($10 \lesssim |\Btheta|_{\mathrm{max}} \lesssim 20$~T/s), moderate-strength ($20 \lesssim |\Btheta|_{\mathrm{max}} \lesssim 40$~T/s) and strong ($|\Btheta|_{\mathrm{max}} \gtrsim 40$~T/s) fishbone activities. Together with the poloidal magnetic fluctuations $\Btheta$, these statistical studies are performed on fluctuating electron temperature $\Te$, which is measured by the ECE radiometer with 12 radial channels covering sight lines from the core to the edge regions of KSTAR plasmas and frequency-filtered within a band of fishbone activities including chirping and saturated frequencies. $\Te$ is also decomposed into the amplitude envelope and the phase components by using the Hilbert transform, which allows us to scrutinize dynamics on short (phase) and long (amplitude envelope) time scales.

We find that the core activity of double-peaked fishbones for both the conditional average (with respect to $\Btheta$) of normalized amplitude envelopes (Figure \ref{fig:summary_condavg_corr}(a)) and the correlation coefficients with $\Btheta$ (Figure \ref{fig:summary_condavg_corr}(c)) does not depend on the fishbone strength as much as the edge activity does (Figure \ref{fig:summary_condavg_corr}(b)), i.e., the core activity does not increase in proportion to the edge activity. (see Section \ref{sec:stat_DPFB}).

Results from probing the phase difference, i.e., $\Delta \phi = \phi(\Te^{\Edge}) - \phi(\Te^{\Core})$, have indicated that the phase of the edge activity leads that of the core activity for most double-peaked fishbones, except for the very weak fishbones where $\Delta \phi$ are observed to be random and inconclusive (see Figure \ref{fig:all_phase_analysis}). The result $\Delta\phi > 0$ appears to be statistically robust in the region where the measurement was made, namely the outer (low-field side) midplane. However, as we noted at the end of Section \ref{sec:phase_analysis} above, the relative phasing is expected to vary and even change sign at other poloidal angles (see Section 5.2.2. of the second paper in this series~\cite{Bierwage2026} for details.)


We believe that our careful and thorough statistical investigations on double-peaked fishbones support the possibility of the edge activity being primary (cause/source) and the core activity being secondary (effect/receiver) as proposed in the introductory section. The results presented here would have to be incorporated into and explained by theories that attempt to tackle questions like those we posed in the introduction: (i) how are double-peaked fishbones destabilized?, and (ii) how are the oscillations at the inner and outer peak synchronized with respect to both amplitude and phase in a differentially rotating plasma? In terms of the drive, we note that the excitation of an energetic-particle-driven instability in the plasma peripheral region has been reported in the Large Helical Device (LHD)~\cite{Du2015, Ida2018}, and we think that this may also be possible in a tokamak, specifically for the present double-peaked fishbones in KSTAR.

Given the importance of the fishbone instability associated with energetic particles for high-performance fusion-grade plasmas in general, as well as various possible challenges and opportunities that could be associated with core-edge coupling phenomena like our double-peaked fishbones in particular, we plan to look into these and related questions as an important future work. Ref~\cite{Bierwage2026} (Part II of this paper series) reports, based on thorough numerical studies, that the core activity does not require a core-localized drive and can be excited from the edge sub-resonantly.

\section*{Acknowledgments}

The authors thank Dr.~M.J.~Choi, Dr.~Jaehyun Lee, and Dr.~M.H.~Kim for providing the ECEI data. The authors also thank Dr.~Panith Adulsiriswad for his assistance when testing alternative statistical causality analysis methods. This work is supported by National R\&D Program through the National Research Foundation of Korea (NRF) funded by the Ministry of Science and ICT (grant numbers RS-2022-00155917 and NRF-2021R1A2C2005654). J.K., K.D.L. and J.G.B. are supported by the R\&D Program of ``KSTAR Experimental Collaboration and Fusion Plasma Research (EN2501-16)'' through the Korea Institute of Fusion Energy (KFE) funded by Korea Ministry of Science and ICT (MSIT).

\appendix

\section{Radially alternating phase structure within the edge activity of strong fishbones}\label{appendix:pi_phase_diff}

\begin{figure}
    [tb]\centering
    \includegraphics[width=0.35\textwidth,keepaspectratio]{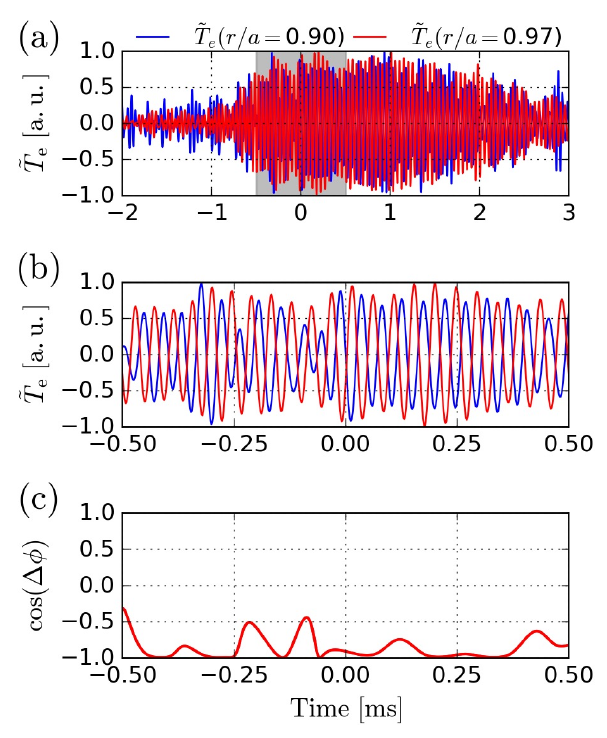}
    \caption{An example of one double-peaked fishbone event viewed by two radially adjacent edge channels, $\Te^{(1)}\equiv\Te(r/a=0.78)$ (blue) and $\Te^{(2)}\equiv\Te(r/a=0.84)$ (red), of the ECE radiometer from KSTAR shot \#22705 for (a) $-2.0 \leq t \leq 2.0$~ms, and (b) $-0.5 \leq t \leq 0.5$~ms, corresponding to the gray shaded region in (a). (c) Temporal evolution of the instantaneous phase difference between $\Te^{(1)}$ and $\Te^{(2)}$.}
    \label{fig:22705_pi_phase_diff}
\end{figure}

We have identified a structure with radially alternating phase between radially adjacent edge channels of the ECE radiometer (see Figure \ref{fig:22705_ECEI_pi_diff}). As an example, Figure \ref{fig:22705_pi_phase_diff}(a) shows two ECE signals during KSTAR shot \#22705, i.e., $\Te^{(1)}\equiv\Te(r/a=0.78)$ in blue and $\Te^{(2)}\equiv\Te(r/a=0.84)$ in red, from a single double-peaked fishbone activity for $-2.0 \leq t \leq 2.0$~ms, where $t=0$ is defined as the time at which the amplitude envelope of $\Btheta$ reaches its maximum. For a clearer view, Figure \ref{fig:22705_pi_phase_diff}(b) shows the same ECE signals for $-0.5 \leq t \leq 0.5$~ms, corresponding to the gray shaded region in Figure \ref{fig:22705_pi_phase_diff}(a). Temporal evolution of the instantaneous phase, that is $\cos(\Delta\phi)$ with $\Delta \phi = \phi(\Te^{(2)})-\phi(\Te^{(1)})$, is also shown in Figure \ref{fig:22705_pi_phase_diff}(c), indicating indeed that $\Delta\phi \sim \pm\pi$.

\begin{figure}
    [tb]\centering
    \includegraphics[width=0.45\textwidth,keepaspectratio]{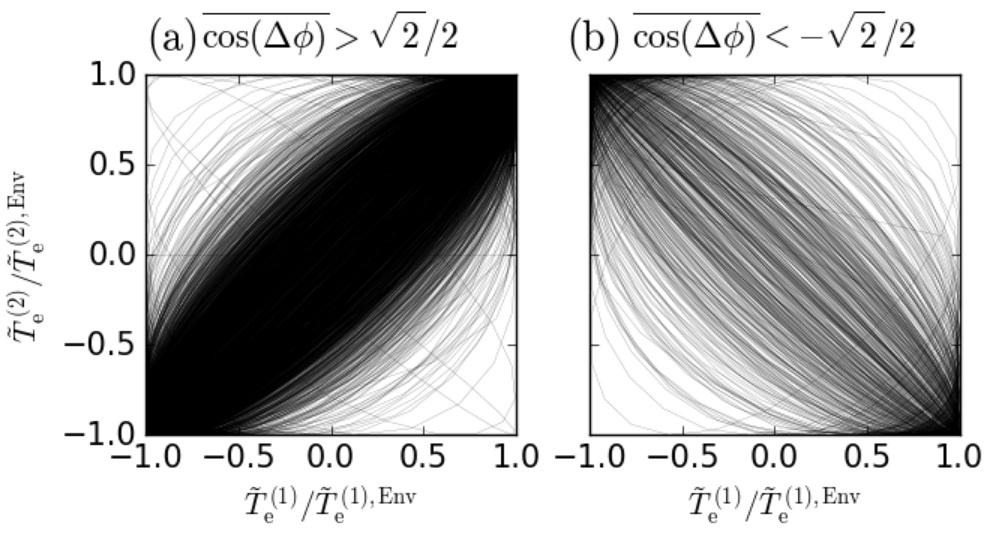}
    \caption{Lissjous curves $L=\left (\frac{\Te^{(1)}}{\Te^{(1),\Env}},\frac{\Te^{(2)}}{\Te^{(2),\Env}} \right )$ for $-0.5\leq t \leq0.5$~ms of double-peaked fishbone activities from KSTAR shot \#22705 for (a) $\overline{\cos{(\Delta\phi)}} > 1/\sqrt(2)$ (unlikely to have the radially alternating phase structure) and (b) $\overline{\cos{(\Delta\phi)}} < 1/\sqrt(2)$ (likely to have the radially alternating phase structure). The radially alternating phase structure occurs only rarely.}
    \label{fig:22705_pi_phase_diff_lissajous}
\end{figure}

\begin{figure}
    [tb]\centering
    \includegraphics[width=0.35\textwidth,keepaspectratio]{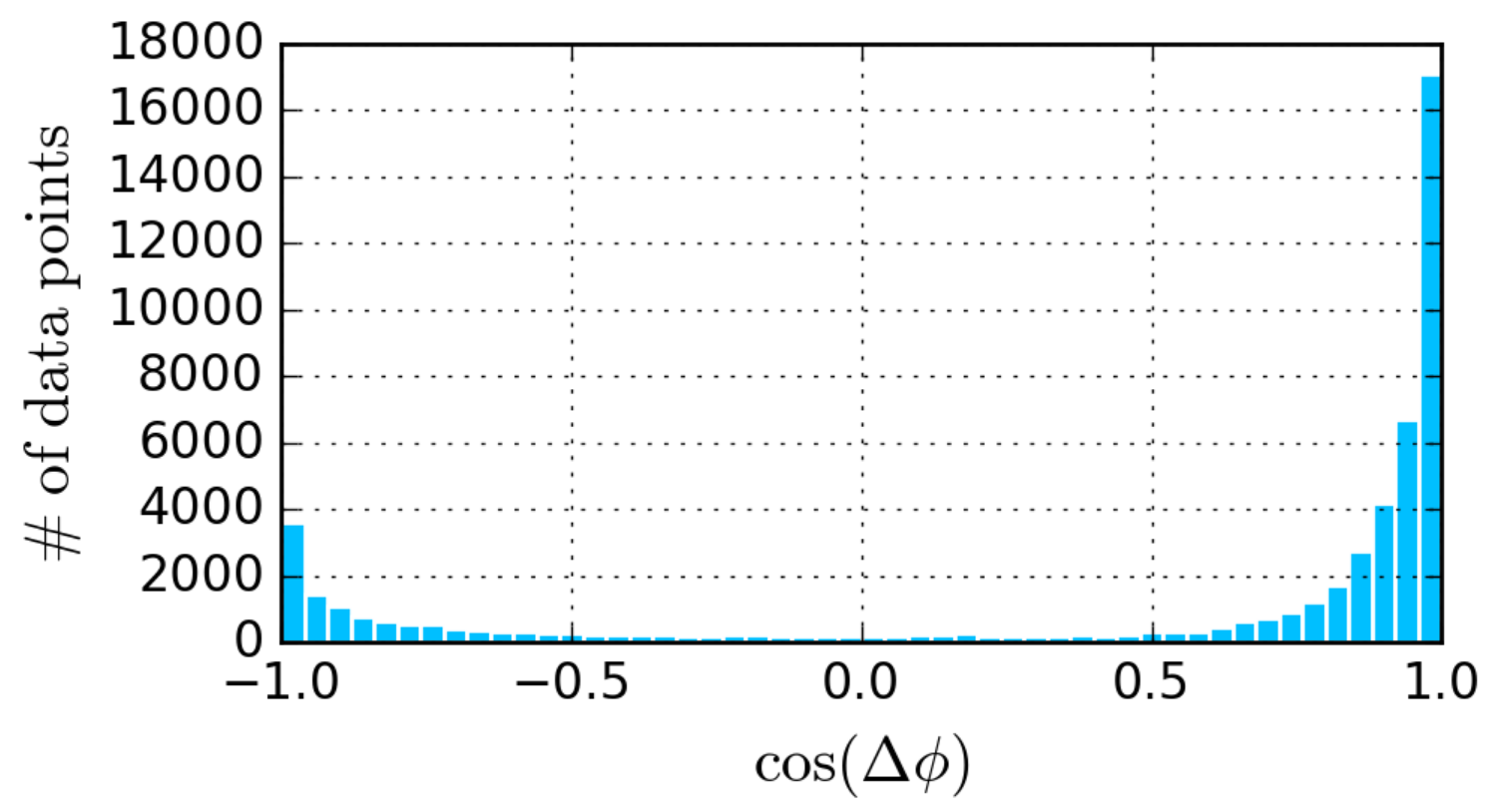}
    \caption{Histogram of $\cos{(\Delta\phi)}$, with $\Delta \phi = \phi(\Te^{(2)})-\phi(\Te^{(1)})$, for $-0.5\leq t \leq0.5$~ms of double-peaked fishbone activities from KSTAR shot \#22705.}
    \label{fig:22705_cos_phi_pi_diff}
\end{figure}

The Lissajous curves are obtained using $\Te^{(1)}$ and $\Te^{(2)}$ from KSTAR shot \#22705 to illustrate how often the radially alternating phase structure occurs in the edge plasmas. To distinguish the radially alternating phase structure, we have averaged an instantaneous phase, i.e., $\cos(\Delta\phi)$, over each event of a fishbone activity for $-0.5\leq t \leq0.5$~ms, which we denote as $\overline{\cos{(\Delta\phi)}}$. Figure \ref{fig:22705_pi_phase_diff_lissajous}(a) and (b) show the Lissajous curves of numerous fishbone activities with $\overline{\cos{(\Delta\phi)}} > 1/\sqrt(2)$ (unlikely to have the radially alternating phase structure) and with $\overline{\cos{(\Delta\phi)}} < 1/\sqrt(2)$ (likely to have the radially alternating phase structure), respectively. Figure \ref{fig:22705_pi_phase_diff_lissajous} shows qualitatively that the radially alternating phase structure occurs infrequently as the corresponding Lissajous curves are much sparser than the other case.

This can also be seen from the histogram of $\cos{(\Delta\phi)}$ in Figure \ref{fig:22705_cos_phi_pi_diff}. The histogram has two peaks, that is one at $\cos(\Delta \phi) = 1$, indicating a small phase difference, and the other at $\cos(\Delta \phi) = -1$, corresponding to the radially alternating phase structure. We observe a considerably smaller peak at $\cos(\Delta \phi) = -1$. This suggest that while most of the edge fishbone activities have a small phase difference, there exists a small number of the edge activities with the radially alternating phase structure. Further investigations on such a structure are left as future work.


\section*{References}
\addcontentsline{toc}{section}{References}

\bibliographystyle{unsrt}
\bibliography{Reference}

\end{document}